\documentclass[aps,prb,twocolumn,floatfix,superscriptaddress,longbibliography]{revtex4-2}
\usepackage{lipsum}
\usepackage{bm}
\usepackage{subcaption}
\usepackage{colordvi}
\usepackage{color}
\usepackage{comment}
\usepackage{amssymb}
\usepackage{babel}
\usepackage[latin1]{inputenc}
\usepackage{amsbsy, amstext, amscd, amsxtra, amsopn, amsfonts, amsthm, amsmath}
\usepackage{hyperref}
\hypersetup{colorlinks=true, linkcolor=blue, citecolor=blue, linktoc=page}
\usepackage{indentfirst}
\usepackage[final]{graphicx}
\graphicspath{ {./images/} }
\usepackage{ragged2e}
\usepackage{titlesec}
\usepackage{cancel}
\usepackage{enumitem}
\usepackage{caption}
\usepackage{breqn}
\usepackage{braket}
\usepackage{multirow}
\usepackage{siunitx}
\usepackage{soul}

\makeatletter
\let\cat@comma@active\@empty
\makeatother

\usepackage[labelfont=bf, justification=Justified, format=plain]{caption}

\setcitestyle{numbers,open={[},close={]}} 

\def\bE{{\overline{E}}}

\newcommand{\bx}{\overline{x}}

\newcommand{\bV}{\overline{V}}

\newcommand{\bU}{\overline{U}}
\newcommand{\bom}{\overline{\Omega}}
\newcommand{\bPhi}{\overline{\Phi}}

\newcommand{\bbE}{\overline{E}}
\newcommand{\cE}{{\cal E}}
\newcommand{\cV}{{\cal V}}
\newcommand{\cH}{{\cal H}}

\newcommand{\cK}{{\cal K}}
\newcommand{\cN}{{\cal N}}
\newcommand{\cO}{{\cal O}}
\newcommand{\cX}{{\cal X}}

\newcommand{\tG}{\widetilde G}

\newcommand{\bd}{\overline{d}}
\newcommand{\bj}{\overline{j}}

\newcommand{\ha}{{\hat a}}

\newcommand{\hn}{{\hat n}}
\newcommand{\hH}{{\hat H}}
\newcommand{\hV}{{\hat V}}

\begin{document}
\title{Nonstandard Hubbard model and electron pairing}

\author{M. Zendra}
\email[Corresponding author: ]{matteo.zendra@unicatt.it}
\affiliation{Dipartimento di Matematica e Fisica and Interdisciplinary Laboratories for Advanced Materials Physics, Universit{\`a} Cattolica del Sacro Cuore, via della Garzetta 48, 25133 Brescia, Italy}
\affiliation{Institute for Theoretical Physics, KU Leuven, Celestijnenlaan 200D, B-3001 Leuven, Belgium}
\affiliation{Istituto Nazionale di Fisica Nucleare, Sezione di Milano, via Celoria 16, I-20133 Milano, Italy}
\author{F. Borgonovi}
\affiliation{Dipartimento di Matematica e Fisica and Interdisciplinary Laboratories for Advanced Materials Physics, Universit{\`a} Cattolica del Sacro Cuore, via della Garzetta 48, 25133 Brescia, Italy}
\affiliation{Istituto Nazionale di Fisica Nucleare, Sezione di Milano, via Celoria 16, I-20133 Milano, Italy}
\author{G. L. Celardo}
\affiliation{Dipartimento di Fisica e Astronomia, Universit{\`a} di Firenze, via Sansone 1, 50019 Sesto Fiorentino, Firenze, Italy}
\affiliation{Istituto Nazionale di Fisica Nucleare, Sezione di Firenze, via Bruno Rossi 1, 50019 Sesto Fiorentino, Firenze, Italy} 
\author{S. Gurvitz}
\email[Corresponding author: ]{shmuel.gurvitz@weizmann.ac.il}
\affiliation{Department of Particle Physics and Astrophysics, Weizmann Institute of Science, 76100 Rehovot, Israel}
\date{\today}
\setlength{\parindent}{20pt}

\begin{abstract}
We present a nonstandard Hubbard model applicable to arbitrary single-particle potential profiles and inter-particle interactions. Our approach involves a novel treatment of Wannier functions, free from the ambiguities of conventional methods and applicable to finite systems without periodicity constraints. To ensure the consistent evaluation of Wannier functions, we develop a perturbative approach, utilizing the barrier penetration coefficient as a perturbation parameter. With the newly defined Wannier functions as a basis, we derive the Hubbard Hamiltonian, revealing the emergence of density-induced and pair tunneling terms alongside standard contributions. Our investigation demonstrates that long-range inter-particle interactions can induce a novel mechanism for repulsive particle pairing. This mechanism relies on the effective suppression of single-particle tunneling due to density-induced tunneling. Contrary to expectations based on the standard Hubbard model, an increase in inter-particle interaction does not lead to an insulating state. Instead, our proposed mechanism implies the coherent motion of correlated electron pairs, similar to bound states within a multi-well system, resistant to decay from single-electron tunneling transitions. These findings carry significant implications for various phenomena, including the formation of flat bands, the emergence of superconductivity in twisted bilayer graphene, and the possibility of a novel metal-insulator transition.
\end{abstract}

\maketitle
\section{INTRODUCTION}
\label{sec:INTRO}
Electron pairing in solids has traditionally been attributed to phonon-mediated attraction. However, a fundamental question is whether repulsive particles can form pairs independently of the presence of phonons. To explore this idea further, we examine two interacting electrons within the same site of a periodic structure, as described by the Hubbard model \cite{hubbard:electrons_corr}. When the on-site two-particle repulsive energy, denoted as $U$, significantly exceeds the tunneling coupling $\Omega$, {\em single-electron} hopping to a neighboring site is strongly suppressed due to the large energy mismatch. Concerning the tunneling of an electron pair, the {\em elastic} two-electron hopping (known as ``co-tunneling'') is also suppressed in the standard Hubbard model. Indeed, the corresponding amplitude representing two consecutive hoppings is a second-order process that involves a large virtual energy variation $\sim 2\Omega^2/U$ \cite{bloch:direct_observation}, which decreases with $U$. Even if weak, this second-order process survives for any finite interaction $U$, so that the repulsive interaction cannot completely localize the electron pair within the framework of the standard Hubbard model.

However, it is evident that the standard Hubbard Hamiltonian fails to capture all the interaction effects \cite{dutta:non_standard, jurgensen:density_induced_optical, jurgensen:observation_density_induced_tunnelling}. For instance, the co-tunneling process can occur with both particles staying together, without changing their total energy, a nonstandard Hubbard process known as pair tunneling (PT) \cite{dutta:non_standard}. Even with increasing $U$, the latter can become a major contributor to the co-tunneling process \cite{gurvitz:twoelectroncorrelated}. Indeed, even in the case of single-electron tunneling coupling suppression $(\Omega\to 0)$, PT remains uninhibited, offering an effective mechanism for electron pairing, independent of the attractive interaction. A similar idea was proposed by P. W. Anderson in the theory of cuprate superconductivity \cite{leggett:bookquantum}. 
\begin{figure*}[t]
\centering
{\includegraphics[width=17.2cm]{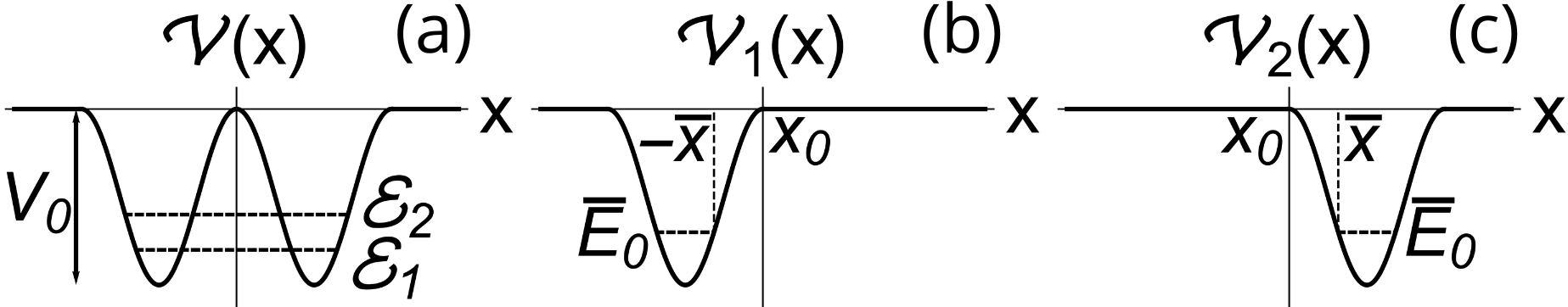}}
\caption{(a) Symmetric double-well potential $\cV(x)=\cV_1(x)+\cV_2(x)$, with lattice depth $V_0$. Dashed lines represent the first two lowest-band energy levels $\cE_{1,2}$. (b,c) Single-well potentials $\cV_{1,2}(x)$, with classical turning points $\mp \,\bx$, so that $\cV_{1,2}(\mp \,{\overline{x}})=\bE_0$, the single-site ground state energy. The separation point $x_0$ is defined so that $\cV_1(x_0)=\cV_2(x_0)=0$.}
\label{fig1}
\end{figure*}
Inspired by this idea, we demonstrate how this specific mechanism can be realized within the framework of a nonstandard Hubbard model, which includes both the pair tunneling term and the density-induced tunneling (DT) term, also known as \textit{bond-charge interaction} \cite{dutta:non_standard, jurgensen:density_induced_optical, jurgensen:observation_density_induced_tunnelling,supercond_state:hirsch, hubbard_nn_bc:strack, ferm_corr_hopp:karnaukhov, eta_pairing:deboer, weak_coupl_hubb:japaridze, two_point_entangl:anfossi, quantum_phase_diagr:dobry, interact_top_ins:montorsi}. 
While the influence of the PT term on superconductivity is rather obvious, the effect of the DT term (with the adequate sign) also favors superconductivity. Indeed, analytical demonstrations have highlighted the role of this term in supporting the emergence of superconductivity within models characterized by repulsive on-site interaction at half-filling \cite{incomm_unconv:aligia}. Thus, the DT and PT terms play a crucial role in electron dynamics, both in terms of their magnitude and sign. Specifically, we show that the DT term, in presence of a long-range inter-particle interaction, has the ability to lower and even totally suppress the single-particle coupling, due to an effective mean-field generated by the other particles, thus providing stability of the electron pair.

An extension of the standard Hubbard model concerning strongly correlated systems has been explored long ago in \cite{hirsch:electron_hole}. However, only recently the nonstandard Hubbard model has attracted more attention, particularly due to novel experimental results with ultracold atoms in optical lattices \cite{jurgensen:observation_density_induced_tunnelling,dutta:non_standard,photon_assisted_tunn:ma, floq_eng_corr_tunn:meinert, enh_sign_change:gorg}, as well as because they have been shown to host many different effects, ranging from
superconducting pairing to localization \cite{knothe2024extended, affleck:heishubb, gilmutdinov:interplayext, chen:superconducting, adebanjo:ubiquity, kundu2023cdmfthfd, wrz:nonmonotonic, frey:hilbertspace, nicokatz:mbl}.
Currently, the accurate evaluation of nonstandard Hubbard terms and the understanding of their influence on the dynamics of correlated systems remain open problems. Indeed, these terms are closely related to the overlap of Wannier functions (WFs) from adjacent sites, often accurately represented by the corresponding orbital wave functions. However, their overlap, crucially dependent on their tails situated in neighboring sites, significantly affects both the magnitude and sign of nonstandard Hubbard terms, on which consensus is yet to be reached \cite{hirsch:electron_hole,luhmann:multiorb_dens_ind,dutta:non_standard}.

In what follows, we present a new approach for evaluating the WFs in a multi-well potential, based on the two-potential approach (TPA) to tunneling problems, originally developed for tunneling to the continuum \cite{gurvitz:decay_width,gurvitz:novel_approach,gurvitz:modified_pot_approach}, which allows for an accurate evaluation of nonstandard Hubbard terms. 
Specifically, after a proper definition of the WFs of a multi-well potential in Sec.~\ref{sec:WF}, we present the TPA in Sec.~\ref{sec:TPA}, and we apply it to the case of a triple-well potential in Sec.~\ref{sec:TW}. Finally, in Sec.~\ref{sec:NSH} we analyze the effect of the PT and DT terms, for both a contact interaction and a long-range constant interaction. Specifically, in Sec.~\ref{subsec:NSH_A} we analyze the simple case of a square double-well potential, showing that the DT term can effectively suppress the total single-particle tunneling amplitude only in presence of a long-range interaction. In Sec.~\ref{subsec:NSH_B}, we study the dynamics of two electrons with parallel spins in a square triple-well potential. In particular, we show under which conditions the nonstandard DT and PT terms become significant, and when the nonstandard Hubbard model should be used instead of the extended Hubbard model (which neglects DT and PT contributions).

\section{WANNIER FUNCTIONS}
\label{sec:WF}
\begin{figure*}[t]
\centering
{\includegraphics[width=17.2cm]{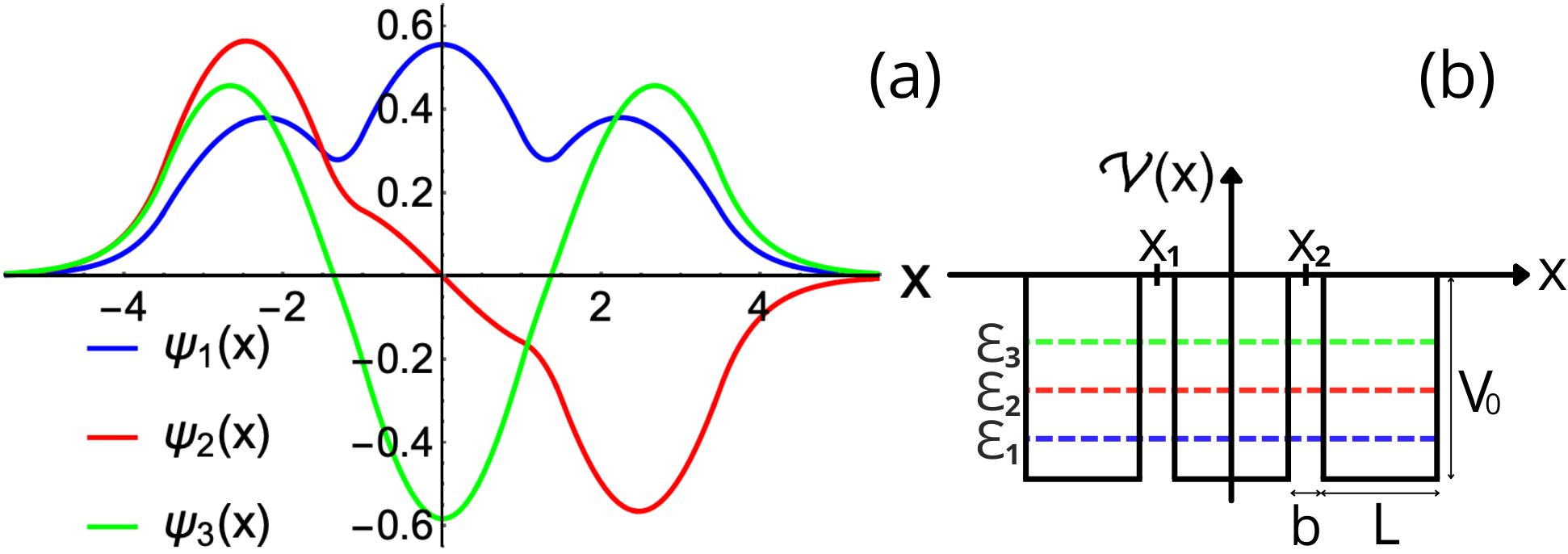}}
\caption{(a) First three exact eigenfunctions $\psi_1(x)$ (blue curve), $\psi_2(x)$ (red curve) and $\psi_3(x)$ (green curve) of a square triple-well potential. (b) Symmetric square triple-well potential. The three wells have width $L$ and depth $V_0$, and are separated by barriers of width $b$, where $x_{1,2}$ are the separation points. Dashed colored lines represent the first three exact energy levels of the system $\cE_{1,2,3}$, corresponding to the eigenfunctions shown in panel (a), which read $\cE_1=-4.171$, $\cE_2=-3.897$ and $\cE_3=-3.545$. Parameters: $L=2$, $b=0.5$ and $V_0=5$, in arbitrary units.}
\label{fig2} 
\end{figure*}
Let us consider a particle placed in an $N$-site potential chain 
\begin{equation}
\cV(x)=\sum_{j=1}^N \cV_j(x)\,,
\end{equation}
where $\cV(x)\to 0$ as $x\to\pm\infty$. The exact eigenstates are obtained from the Schr\"odinger equation (we take $\hbar=1$)
\begin{equation}
\cH\ket{\psi_k}\equiv \left(-\frac{\nabla_x^2} {2m}+\cV(x)\right)\ket{\psi_k}=\cE_k\ket{\psi_k}\,,
\label{eq:ham}
\end{equation}
with boundary conditions at infinity ($x\to\pm\infty$) given by
\begin{equation}
\psi_k(x)\sim e^{-\sqrt{-2m\cE_k}|x|}\,,
\label{eq:bc}
\end{equation}
that uniquely define the bound state energy spectrum $(\cE_k<0)$ of the exact Hamiltonian $\cH$. We assume that the $N$ lowest eigenstates form a band, well separated from the other eigenstates of the spectrum. 

We consider the corresponding tight-binding tunneling Hamiltonian $H_N$, which describes the lowest band of the exact Hamiltonian $\cH$, given by
\begin{equation}
H_N=\sum_{j=1}^{N}\bbE_j\ket{\Psi_j}\bra{\Psi_j}+\sum_{j=1}^{N-1}\bom_j\left(\ket{\Psi_j}\bra{\Psi_{j+1}}+H.c.\right)\,,
\label{tunn_ham}
\end{equation}
where $\bE_j$ represents the single-site energy, $\bom_j$ is the nearest neighbor tunneling coupling, while $\Psi_j(x)=\braket{x|\Psi_j}$ are the WFs. In order to define $\bE_j$ and $\bom_j$ in a consistent way, we identify the spectrum of the tunneling Hamiltonian in Eq.~\eqref{tunn_ham} with the one of the lowest band of the original Hamiltonian $\cH$ in Eq.~\eqref{eq:ham}. When employing such a procedure, unlike when solving exactly the Schr\"odinger Eq.~\eqref{eq:ham}, we are neglecting the influence of inter-band transitions on electrons' motion. Indeed, if the lowest band is sufficiently separated from the other bands, the exact spectrum obtained from Eq.~\eqref{eq:ham} and the one of the tunneling Hamiltonian will produce the same dynamics. Therefore, we diagonalize the Hamiltonian $H_N$ by a unitary transformation $R$ and then we apply the same transformation to the lowest-band spectrum, namely $\cE_k$ and $\ket{\psi_k}$, to obtain the WFs. In particular, these are uniquely defined by 
\begin{equation}
\ket{\Psi_k}=\sum_{k'=1}^N R_{k k'}\ket{\psi_{k'}}\,.
\label{untr}
\end{equation}
Notice that the exact eigenfunctions $\psi_k(x)$ do not contain uncertainty, as belonging to the bound-state spectrum of the Schr\"odinger Eq.~\eqref{eq:ham}. In the following, we will illustrate the unitary transformation given in Eq.~\eqref{untr} for the double-well and triple-well potential cases (for details, see the Supplemental Material \cite{supplmat}).

Let us exemplify this method considering the symmetric double-well potential $\cV(x)$ in Fig.~\ref{fig1} (a), where the lowest band contains two eigenstates $\psi_{1,2}(x)$, with corresponding eigenenergies $\cE_{1,2}$. The tunneling Hamiltonian of this system is given by Eq.~\eqref{tunn_ham} for $N=2$, and can be explicitly written as 
\begin{equation}
H_2=\bE_0\sum_{j=1}^2\ket{\Psi_j}\bra{\Psi_j}+
\bom_0\left(\ket{\Psi_1}\bra{\Psi_{2}}+H.c.\right)\,.
\end{equation}
By diagonalizing $H_2$ through the unitary transformation $R$ in Eq.~\eqref{untr}, and identifying its eigenspectrum with $\cE_{1,2}$ and $\psi_{1,2}(x)$, we find 
\begin{subequations}
\label{Wannier_functions}
\begin{align}
&\bbE_0=\frac{1}{2}\left(\cE_1+\cE_2\right)\,, \quad \bom_0= \frac{1}{2}\left(\cE_1-\cE_2\right)\,,
\label{Wannier_functionsa}\\
&\Psi_{1,2}(x)=\frac{1}{\sqrt{2}}\Big[\psi_1(x)\pm \psi_2(x)\Big]\,.
\label{Wannier_functionsb}
\end{align}
\end{subequations}
In contrast with the ``extended'' eigenstates $\psi_{1,2}(x)$, the WFs $\Psi_{1,2}(x)$ are localized respectively in the left and right well, although their tails are extended to the neighboring wells. 

This procedure can be easily extended for the symmetric triple-well potential $(N=3)$ in Fig.~\ref{fig2}, where the lowest band consists of three eigenstates $\psi_{1,2,3}(x)$ with energies $\cE_1<\cE_2<\cE_3$. The corresponding tight-binding tunneling Hamiltonian, given by Eq.~\eqref{tunn_ham} for $N=3$, can be explicitly written as 
\begin{equation}
H_3=\bE_0\sum_{j=1}^3\ket{\Psi_j}\bra{\Psi_j}+
\bom_0\left(\ket{\Psi_1}\bra{\Psi_{2}}+\ket{\Psi_2}\bra{\Psi_{3}}+H.c.\right)\,.
\label{tunn_ham3}
\end{equation}
Following the same procedure, i.e. by diagonalizing $H_3$ and identifying the obtained spectrum with the exact lowest-band one, we obtain 
\begin{equation}
\label{wf3}
\begin{aligned}
&\bE_0=\frac{1}{2}\left(\cE_1+\cE_3\right)\,,
\quad\bom_0=\frac{1}{2\sqrt{2}}\left(\cE_1-\cE_3\right)\,,\\
&\Psi_{1}(x)=\frac{1}{2}\psi_1(x)+\frac{1}{\sqrt{2}}\psi_2(x)
+\frac{1}{2}\psi_3(x)\,,\\
&\Psi_{2}(x)=\frac{1}{\sqrt{2}}\Big[\psi_1(x)
-\psi_3(x)\Big]\,,\\
&\Psi_{3}(x)=\frac{1}{2}\psi_1(x)-\frac{1}{\sqrt{2}}\psi_2(x)
+\frac{1}{2}\psi_3(x)\,.
\end{aligned}
\end{equation}
As in the previous case, the WFs $\Psi_{1,2,3}(x)$ are respectively localized in the left, middle and right well, and are uniquely defined. Let us point out that our approach for a consistent determination of the tunneling Hamiltonian parameters and the related WFs can be generalized for an arbitrary number of potential wells $N$, regardless the periodicity of $\cV(x)$. Additionally, we observe that for a {\em periodic} potential $\cV(x)$, in the limit $N\to\infty$, this procedure looks similar to the method used to derive a set of localized WFs from the Bloch functions, subjected to periodic boundary conditions, through a unitary transformation. However, due to the additional ``Gauge freedom'', the resulting WFs become strongly nonunique, so that different choices of the gauge correspond to different sets of WFs having different shapes and spreads. A widely used approach to avoid the Gauge freedom consists in a proper choice of the unitary transformation of the Bloch functions that enforces the maximal localization of the WFs (see \cite{marzari:wannier} for a detailed discussion). However, this procedure does not guarantee that the tunneling Hamiltonian dynamics corresponds to that obtained from the exact solution of the original multi-well Schr\"odinger equation. 

In contrast, our approach is based on this correspondence, which allows to uniquely construct the tunneling Hamiltonian and the WFs by assuming only the single-band approximation. Notice that the resulting WFs, although localized at the corresponding site, exhibit \textit{tails} penetrating to neighboring sites. These tails play a crucial role in the evaluation of the nonstandard Hubbard terms, as we will show in the following. On the contrary, the condition of maximal localization of the WFs would decrease correspondingly the contribution from these tails, and therefore the amplitude of the nonstandard Hubbard terms. Since our approach relates the WFs to the exact Schr\"odinger eigenstates, in the next section we present a consistent perturbative approach for their evaluation in terms of single-site orbitals. 

\section{TWO-POTENTIAL APPROACH}
\label{sec:TPA}
Let us consider the symmetric double-well potential in Fig.~\ref{fig1} (a), given by the sum of two single-well potentials, $\cV(x)=\cV_1(x)+\cV_2(x)$, such that $\cV_1(x)=0$ for $x\ge x_0$ and $\cV_2(x)=0$ for $x\le x_0$, where $x_0=0$ is the separation point, see Fig.~\ref{fig1} (b,c). The lowest eigenstates (orbitals) of the left- and right-well Hamiltonians are obtained from 
\begin{equation}
\left(-\frac{\nabla_x^2} {2m}+\cV_{1,2}(x)\right)\Phi_0^{(1,2)}(x)=E_0\Phi_0^{(1,2)}(x)\,,
\label{orb}
\end{equation}
with the following boundary conditions:
\begin{equation}
\label{eq:bc2}
\begin{aligned}
&\Phi_0^{(1)}(x)\sim e^{\sqrt{-2mE_0}x} \quad {\rm as} \quad x\to -\infty\,,\\
&\Phi_0^{(1)}(x)=\Phi_0^{(1)}(0)e^{-\sqrt{-2mE_0}x} \quad {\rm as}\quad 
x\ge x_0\,,
\end{aligned}
\end{equation}
and similarly for $\Phi_0^{(2)}(x)=\Phi_0^{(1)}(-x)$. These orbitals can be used as a basis to obtain the eigenstates $\psi_{1,2}(x)$ and the WFs $\Psi_{1,2}(x)\equiv\Psi_{L,R}(x)$, through a perturbative approach. For instance, we could consider the left-well orbital $\Phi_0^{(1)}(x)$ as the unperturbed state and the right-well potential $\cV_2(x)$ as the perturbation (or vice versa).

However, such perturbative approach does not include a small parameter, which makes the corresponding expansion unusable. This issue can be solved by employing the TPA, which uses an alternative expansion in powers of the orbitals overlap $\beta \equiv \braket{\Phi_0^{(1)} |\Phi_0^{(2)}}$, a small parameter proportional to the barrier penetration coefficient
\begin{equation}
T_0=\exp \left(-\int\limits_{-\bx}^{\bx} |p(x')|\,dx'\right)\ll 1\,.
\label{T0_barr_coeff}
\end{equation}
Here, $p(x)$ represents the (imaginary) momentum under the potential barrier, and $\pm\bx$ are the classical turning points, shown in Fig.~\ref{fig1} (b,c) (for details, see the Supplemental Material \cite{supplmat}). Using this approach, we derive the tunneling Hamiltonian parameters in Eq.~\eqref{Wannier_functionsa}, which read
$$\bE_0=E_0+\cO\left(\beta^2\right)\,,$$
$$\bom_0=\Omega_0+\cO\left(\beta^2\right)\,,$$
where $E_0$ is given by Eq.~\eqref{orb}, and 
\begin{equation}
\label{eq:barde}
\Omega_0= -\sqrt{\frac{2|E_0|}{m}}\Big[\Phi_0(0)\Big]^2\propto T_0
\end{equation}
is a simplified (1D) version of the well-known Bardeen formula \cite{bardeen:tunnelling_many_body}. Similarly, we obtain
$$\cE_{1,2}=E_\pm +\cO \left(\beta^2\right)\,,$$
where $E_\pm=E_0\pm\Omega_0$. Consequently, all the parameters of the tunneling Hamiltonian are completely determined by the single-well orbitals. At first glance, we may expect to derive the eigenstates $\psi_{1,2}(x)\equiv\psi_{1,2}(E_\pm,x)$ from Eq.~\eqref{Wannier_functionsb} by replacing the WFs $\Psi_{1,2}(x)$ with the corresponding orbitals $\Phi_0^{(1,2)}(x)\equiv\Phi_0^{(1,2)}(E_0,x)$ given by Eq.~\eqref{orb}, so that
\begin{equation}
\psi_{1,2}(E_\pm,x) \simeq \frac{1}{\sqrt{2}}\left[\Phi_0^{(1)}(E_0,x)\pm\Phi_0^{(2)}(E_0,x)\right]\,.
\label{eigenn}
\end{equation}
However, Eq.~\eqref{eigenn} exhibits an inconsistency between the energy arguments of $\psi_{1,2}(E_\pm,x)$ and $\Phi_0^{(1,2)}(E_0,x)$. To solve this issue, we introduce an energy shift in the orbital functions by replacing the ground state energy $E_0$ with a free parameter $E<0$. The resulting modified orbitals $\bPhi^{(1,2)}(E,x)$ (normalized to unity) are obtained from Eq.~\eqref{orb} with the substitution $E_0\to E$ and imposing the boundary condition at infinity given in Eqs.~\eqref{eq:bc2}. However, unlike $\Phi_0^{(1,2)}(E_0,x)$, the modified orbitals $\bPhi^{(1,2)}(E,x)$ are defined respectively on two different segments 
$$\cX_1=(-\infty, 0) \quad {\rm and} \quad \cX_2=(0,\infty)\,,$$
and vanish elsewhere. As a result, they are {\em non-overlapping}, and therefore {\em orthogonal}. 
Replacing $\Phi_0^{(1,2)}(E_0,x)$ in Eq.~\eqref{eigenn} with $\bPhi^{(1,2)}(E_\pm,x)$, we obtain 
\begin{equation}
\psi_{1,2}(E_\pm,x)=\frac{1}{\sqrt{2}}\left[\bPhi^{(1)}(E_\pm,x)
\pm\bPhi^{(2)}(E_\pm,x)\right]\,,
\label{eigen_site}
\end{equation}
which gives the {\em exact} result for $\psi_{1,2}(E_\pm,x)$, in contrast with Eq.~\eqref{eigenn}. Indeed, the exact treatment of the Schr\"odinger Eq.~\eqref{eq:ham} involves solving it on the two segments and combining the results by imposing the continuity condition at the separation point. This condition is automatically satisfied if $E_\pm$ are the energies of the symmetric and anti-symmetric states, respectively. 

Substituting Eq.~\eqref{eigen_site} into Eq.~\eqref{Wannier_functionsb}, we obtain the exact left- and right-well WFs, $\Psi_{L,R}(x)$, in terms of the modified orbitals:
\begin{equation}
\begin{aligned}
&\Psi_{L}(x)=\frac{1}{2}\left[\bPhi_+^{(1)}(x)
+\bPhi_+^{(2)}(x)+\bPhi_-^{(1)}(x)-\bPhi_-^{(2)}(x)\right]\,,\\
&\Psi_{R}(x)=\frac{1}{2}\left[\bPhi_+^{(1)}(x)
+\bPhi_+^{(2)}(x)-\bPhi_-^{(1)}(x)+\bPhi_-^{(2)}(x)\right]\,,
\end{aligned}
\label{wan_} 
\end{equation}
where $\bPhi_\pm^{(1,2)}(x)\equiv \bPhi^{(1,2)}(E_0\pm\Omega_0, x)$. Expanding the modified orbitals in powers of $\Omega_0$ and neglecting $\cO\left(\Omega_0^2\right)$ terms (since $\Omega_0\propto \beta\propto T_0$) we obtain
\begin{equation}
\bPhi_\pm^{(1,2)}(x) = \bPhi_0^{(1,2)}(x)\pm \Omega_0\,\partial_E\bPhi^{(1,2)}_0(x)\,,
\label{expL}
\end{equation}
where 
\begin{equation}
\bPhi_0^{(1,2)}(x)\equiv
\begin{cases}
\Phi_0^{(1,2)}(E_0,x) \quad &{\rm for}\,\,x\in\cX_{1,2}\\
0&{\rm elsewhere}
\end{cases}\,,
\label{reduced_orb}
\end{equation}
and
$$\partial_E\bPhi_0^{(1,2)}(x) \equiv \left(\frac{\partial\bPhi^{(1,2)}(E,x)}{\partial E}\right)_{E= E_0}\,.$$
Substituting Eq.~\eqref{expL} into Eqs.~\eqref{wan_}, we get 
\begin{equation}
\begin{aligned}
&\Psi_{L}(x)=\bPhi_0^{(1)}(x) 
+\Omega_0\partial_E\bPhi_0^{(2)}(x)\,,\\
&\Psi_{R}(x)=\bPhi_0^{(2)}(x) 
+\Omega_0\partial_E\bPhi_0^{(1)}(x)\,,
\end{aligned}
\label{wan_approx}
\end{equation}
which represents our main result for the WFs. Looking at Eqs.~\eqref{wan_approx}, we can observe that each WF consists of two {\em non-overlapping} terms, describing respectively the WF inside the respective well (first term) and its tail penetrating into the neighboring well (second term), which is $\propto\Omega_0$ and therefore much smaller than the first term. Since $\bPhi^{(1,2)}(E,x)$ are normalized to unity for any $E$, we can explicitly demonstrate the orthogonality of the WFs by using 
$$\partial_E\int\limits_{-\infty}^0 \left[\bPhi^{(1)}(E,x)\right]^2\,dx=0\,,$$
so that
\begin{equation}
\braket{\Psi_L|\Psi_R}=
2\Omega_0\int\limits_{-\infty}^0\bPhi^{(1)}_0(x)\partial_E\bPhi^{(1)}_0(x)\,dx=0\,.
\label{orth}
\end{equation}
Eq.~\eqref{orth} represents the overlap of the orbital $\bPhi_0^{(1)}(x)$, which is nodeless, with the tail of the WF belonging to the adjacent well , see Eqs.~\eqref{wan_approx}. From Eq.~\eqref{orth}, it clearly follows that the WF tail must change its sign, deeply affecting the amplitudes of the nonstandard Hubbard terms. Finally, we point out that Eqs.~\eqref{wan_approx} are valid for an arbitrary multi-well system. In the next section, we exemplify this by comparing the WFs given by Eqs.~\eqref{wan_approx} with the exact numerical results for a symmetric square triple-well potential.

\section{TWO-POTENTIAL APPROACH FOR A TRIPLE-WELL POTENTIAL}
\label{sec:TW}
In this section, we explicitly demonstrate the accuracy of our analytical approach, by analyzing the WFs of the symmetric square triple-well potential shown in Fig.~\ref{fig2} (b). Specifically, we evaluate the WFs by using the TPA and we compare them with the exact WFs given by Eqs.~\eqref{wf3}, as well as with the corresponding orbital functions. For simplicity, we consider a square well potential, since its shape allows us to obtain simple analytical expressions for the WFs, which will be used for the evaluation of the nonstandard Hubbard terms, highlighting their explicit dependence on the quantum well parameters. 

The triple-well spectrum, namely the eigenfunctions $\psi_k(x)\equiv\psi_k(\cE_k,x)$ and the eigenvalues $\cE_k$, is obtained by solving the Schr\"odinger Eq.~\eqref{eq:ham} with boundary conditions given by Eq.~\eqref{eq:bc}. We focus on the three lowest-band eigenstates (with $k=1,2,3$) displayed in Fig.~\ref{fig2} (a). The corresponding exact left-, middle- and right-well WFs $\Psi_{L,M,R}(x)$ can be obtained from the lowest-band eigenstates through Eqs.~\eqref{wf3}. On the other hand, we notice that the energy $\bE_0$ in the tunneling Hamiltonian in Eq.~\eqref{tunn_ham3} corresponds to the energy of the lowest orbital $\Phi_0(x)$ given by Eq.~\eqref{orb}, by considering the single-well potential $$\cV(x)=-V_0 \quad {\rm for}\quad -\frac{L}{2}<x<\frac{L}{2}\,.$$
Specifically, the lowest single-well orbital can be written as 
\begin{align}
\Phi_0(x)=\cN_0
\begin{cases}
\sqrt{1-\frac{|E_0|}{V_0}}e^{q_0 \left(x +\frac{L}{2}\right)}
\,&{\rm for}\,-\infty <x<-\frac{L}{2}\\
\cos \left(p_0 x\right)\,&{\rm for}\,-\frac{L}{2}<x<\frac{L}{2}\\
\sqrt{1-\frac{|E_0|}{V_0}}e^{-q_0 \left(x -\frac{L}{2}\right)}\,&{\rm for}\,\quad\frac{L}{2}<x<\infty
\end{cases}\,,
\label{sqorb}
\end{align} 
where $p_0=\sqrt{2m(V_0+E_0)}$, $q_0=\sqrt{-2mE_0}$ and $\cN_0=\sqrt{2q_0/(2+L q_0)}$ is the normalization factor. As a result, the orbital functions for the triple-well system (respectively for the left, middle and right well) read
\begin{equation}
\begin{aligned}
&\Phi_0^{(1)}(x)\equiv \Phi_0(x+L+b)\,,\\
&\Phi_0^{(2)}(x)\equiv \Phi_0(x)\,,\\
&\Phi_0^{(3)}(x)\equiv \Phi_0(x-L-b)\,.
\end{aligned}
\label{triplewell_orb}
\end{equation}
Substituting $\Phi_0\left(\frac{L+b}{2}\right)$ into Eq.~\eqref{eq:barde}, we obtain for the tunneling energy 
\begin{equation}
\Omega_0=-\sqrt{\frac{2|E_0|}{m}}\cN_0^2\left(1-\frac{|E_0|}{V_0}\right)e^{-q_0b}\,.
\label{sqbard}
\end{equation}
For the single-well parameters used in Fig.~\ref{fig2}, solving Eq.~\eqref{orb} and Eq.~\eqref{sqbard} we obtain $E_0=-3.8525$ and $\Omega_0=-0.2216$. These values can be compared with those obtained from the exact numerical solution of the Schr\"odinger equation for the triple-well potential, namely $\bE_0=-3.858$ and $\bom_0=-0.2215$. Their closeness confirms the high accuracy of the TPA for a consistent determination of the tunneling Hamiltonian parameters. 
\begin{figure}[t]
\centering
{\includegraphics[width=8.6cm]{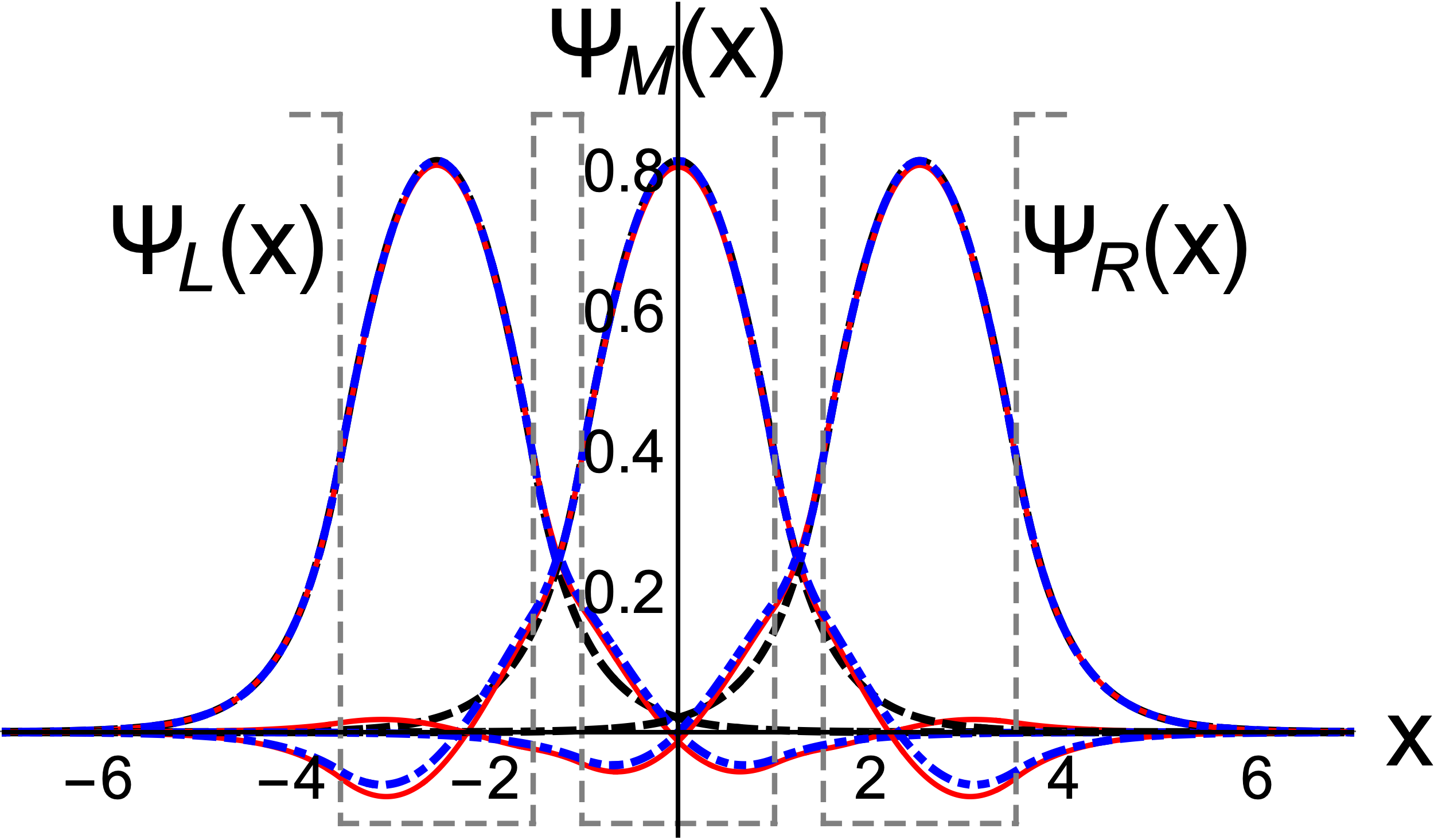}}
\caption{Left-, middle- and right-well WFs for the square triple-well potential shown with dashed grey lines. Red solid curves correspond to exact calculations in Eq.~\eqref{wf3}, blue dashed curves show our analytical results obtain with the TPA in Eqs.~\eqref{wan3}, and black dashed curves show the orbital functions $\Phi_0^{(1,2,3)}(x)$ in Eqs.~\eqref{triplewell_orb}. Parameters: $L=2$, $b=0.5$ and $V_0=5$, in arbitrary units.}
\label{fig3} 
\end{figure}

Finally, let us evaluate the corresponding WFs, that can be obtained by extending Eqs.~\eqref{wan_approx} to a triple-well system. By following the same procedure of the square double-well potential case, we construct the eigenstates $\psi_k(x)$ through the modified orbitals, with energy shift $\cE_k-E_0\propto\Omega_0$. We then obtain the WFs from the eigenstates $\psi_k(x)$ via the unitary transformation in Eq.~\eqref{untr}. By expanding the resulting WFs in powers of $\Omega_0$ up to $\cO\left(\Omega_0^2\right)$ terms, we get a simple result representing the straightforward extension of Eqs.~\eqref{wan_approx}, given by 
\begin{equation}
\label{wan3}
\begin{aligned}
&\Psi_{L}(x)= \bPhi^{(1)}_0(x) 
+\Omega_0\partial_E\bPhi^{(2)}_0(x)\,,\\
&\Psi_{M}(x)= \bPhi^{(2)}_0(x) 
+\Omega_0\left[\partial_E \bPhi^{(1)}_0(x)+\partial_E \bPhi^{(3)}_0(x)\right]\,,\\
&\Psi_{R}(x)= \bPhi^{(3)}_0(x) 
+\Omega_0\partial_E\bPhi^{(2)}_0(x)\,.
\end{aligned}
\end{equation}
As in the double-well case, $\bPhi^{(1,2,3)}_0(x)$ denote the left-, middle- and right-well modified orbitals, respectively coinciding with $\Phi_0^{(1,2,3)}(x)$ of Eqs.~\eqref{triplewell_orb} on the intervals $(-\infty, x_1)$, $(x_1, x_2)$ and $(x_2, \infty)$, and vanishing elsewhere. The separation points $x_{1,2}$ are taken at the center of the inter-well barriers, as shown in Fig.~\ref{fig2} (b). 

Looking at Eqs.~\eqref{wan3}, we notice that the WFs for the triple-well system are given by the same expressions of the double-well system in Eq.~\eqref{wan_approx}. Indeed, the first term representing the WF inside the respective well is given by the orbital, while the second term (with derivatives) describing the WF tails penetrating to neighboring wells is proportional to $\Omega_0$. Let us remark that the latter represents the energy shift (tunneling energy) for the {\em double-well} potential. Remarkably, even if the energy shift in the triple-well case is different ($\cE_1-E_0=\sqrt{2}\Omega_0$), see Eqs.~\eqref{wf3}, the $\sqrt{2}$ factor cancels out during the derivation, confirming that the WF tail is always determined by the tunneling coupling to neighboring well. A detailed derivation for a generic multi-well system will be given in a separate work.

In Fig.~\ref{fig3}, we compare the WFs $\Psi_{L,M,R}(x)$ in Eqs.~\eqref{wan3} obtained with the TPA (blue dashed curves) with the orbital functions $\Phi_0^{(1,2,3)}(x)$ in Eqs.~\eqref{triplewell_orb} (black dashed curves) and the exact results in Eq.~\eqref{wf3} obtained via numerical calculations (red solid curves). We observe that the orbitals $\Phi_0^{(1,2,3)}(x)$ provide a close approximation to the corresponding exact WFs $\Psi_{L,M,R}(x)$ within each well, despite notable differences in their tails into neighboring wells. Furthermore, the approximate results closely match the exact ones, even in the regions of the tails (beyond the respective well), underscoring the precision of the TPA. Ultimately, we notice that the tails of the WFs into the neighboring wells are less pronounced for the left and right wells compared to the middle well, due to the slightly different boundary conditions for the modified orbitals of the external wells, as described in Eqs.~\eqref{eq:bc2}. 

In the next section, we derive the nonstandard Hubbard terms using our analytical expression for the double-well WFs in Eqs.~\eqref{wan_approx}, and we show how these nonstandard Hubbard terms can be used to suppress single-particle tunneling in presence of long-range inter-particle interaction.

\section{NONSTANDARD HUBBARD HAMILTONIAN}
\label{sec:NSH}
 
\subsection{Distinguishable interacting particles in a symmetric double-well potential}
\label{subsec:NSH_A}
The interaction between two particles in a double-well potential can be described by a two-body {\em repulsive} potential $V(x-y)>0$. Since the many-body basis for two distinguishable particles is given by the tensor product of the single-particle WFs, the matrix elements of the interaction term for two distinguishable particles in the tunneling Hamiltonian basis are given by
\begin{equation}
V_{i'j'ij}=\int \Psi_{i'}(x)\Psi_{j'}(y)V(x-y)\Psi_{i}(x)\Psi_{j}(y)\,dx\,dy\,.
\label{V_int_terms}
\end{equation}
Here, $\Psi_i(x)$ is the WF at site $i=L,R$ of the symmetric double-well potential in Fig.~\ref{fig1} (a). The interaction potential in Eq.~\eqref{V_int_terms} can be decomposed into standard and nonstandard Hubbard terms, corresponding respectively to diagonal $(ij=i'j')$ and off-diagonal $(ij\not =i'j')$ matrix elements. The Hubbard terms can be further separated into the standard Hubbard on-site interaction term $V_{iiii}\equiv U$ (for $i=j$) and the \textit{extended} Hubbard term $V_{ijij}\equiv \bU$ (for $i\not=j$) \cite{dutta:non_standard}, respectively defined as
\begin{subequations}
\label{UandUbar}
\begin{align}
&U=\int\Psi_L^2(x)V(x-y)\Psi_L^2(y)\,dx\,dy\,,
\label{u}\\
&\bU=\int\Psi_L^2(x)V(x-y)\Psi_R^2(y)\,dx\,dy\,.
\label{uu}
\end{align}
\end{subequations}
Similarly, the nonstandard Hubbard terms can be separated into the DT $\left(\Omega_1\right)$ and PT $\left(\Omega_2\right)$ terms, with amplitudes respectively given by
\begin{subequations}
\label{Om1Om2}
\begin{align}
&\Omega_1=\int \Psi_L^2(x)\Psi_L(y)V(x-y)\Psi_R(y)\,dx\,dy\,,
\label{Om1}\\
&\Omega_2=\int \Psi_L(x)\Psi_L(y)V(x-y)
\Psi_R(x)\Psi_R(y)\,dx\,dy\,.
\label{Om2}
\end{align}
\end{subequations}
The physical interpretation of these terms is evident: the DT term $\left(\Omega_1\right)$ represents a single-particle hopping (e.g. $\Psi_{LL}\to\Psi_{LR}$) caused by the interaction with the non-tunneling particle, while the PT term $\left(\Omega_2\right)$ describes the direct (e.g. $\Psi_{LL}\to\Psi_{RR}$) and exchange (e.g. $\Psi_{LR}\to\Psi_{RL}$) two-particle hopping. 
In a double-well potential, the DT term in Eq.~\eqref{Om1} can always be added to the single-particle tunneling, resulting in an effective tunneling $\Omega_{eff}\equiv\Omega_0+\Omega_1$ \cite{dutta:non_standard, jurgensen:observation_density_induced_tunnelling}. Therefore, in principle, the effective tunneling can be suppressed by the interaction when $\Omega_1=-\Omega_0$.

For a repulsive {\it contact} interaction described by
\begin{equation}
V(x-y)=V_\delta \,\delta(x-y)>0\,,
\label{delta}
\end{equation}
the DT and PT terms can be evaluated directly by substituting Eqs.~\eqref{wan_approx} into Eqs.~\eqref{Om1Om2}, obtaining
\begin{subequations}
\begin{align}
&\Omega_1=\Omega_0 V_\delta\int\limits_{-\infty}^0
\left[\bPhi^{(1)}_0(x)\right]^3\partial_E\bPhi^{(1)}_0(x)\,dx\,,
\label{om1}\\
&\Omega_2=2\Omega_0^2 V_\delta\int\limits_{-\infty}^0
\left[\bPhi^{(1)}_0(x)\partial_E\bPhi^{(1)}_0(x)\right]^2\,dx\,.
\label{om2}
\end{align}
\end{subequations}
As expected, the DT term is proportional to $\Omega_0$, while the PT term is proportional to $\Omega_0^2$. From Eq.~\eqref{om1}, we notice that if $\Omega_1/\Omega_0<0$, the effective tunneling coupling $\Omega_{eff}$ could be suppressed by a sufficiently large $V_\delta$. However, comparing Eq.~\eqref{orth} with Eq.~\eqref{om1}, we can see that this suppression cannot occur for a contact interaction. Although this can be easily checked numerically, in the following we show how this results can be obtained by a careful analysis of Eq.~\eqref{om1}. Firstly, let us notice that the difference between Eq.~\eqref{om1} and the orthogonality expressed in Eq.~\eqref{orth} lies in the third power of the orbital function $\left[\bPhi_0^{(1)}(x)\right]^3$. In the latter case, the orbital function $\bPhi_0^{(1)}(0)>0$, while the WF tail $\Omega_0 \partial_E \bPhi_0^{(1)}(x)$ changes its sign inside the integral. Since the integral of their product should be zero, both contributions should cancel each other out. On the other hand, the negative contribution to the integral in Eq.~\eqref{om1} is amplified compared to the positive one, because the value of the orbital $\bPhi_0^{(1)}(x)$ decreases as $x\to 0$, where the WF tail is positive. This implies that $\Omega_1<0$, so that the DT term has always the same sign as $\Omega_0$, and consequently it can only increase the effective single-particle tunneling $|\Omega_{eff}|$.

This outcome undergoes a significant transformation when considering instead of a contact interaction a {\em long-range} one
\begin{align}
V(x-y)=
\begin{cases}
\bV \quad &{\rm for}\quad |x-y|<\bd\\
0 \quad &{\rm elsewhere}
\end{cases}\,,
\label{eq:long}
\end{align} 
where $\bd$ denotes the interaction range. For simplicity, in the subsequent discussion we exclusively focus on this toy-model interaction, even if similar results can be obtained using a more physically realistic \textit{screened Coulomb} interaction, as in \cite{vu:moire_mott}. Moreover, this toy-model allows us to study the general behavior of the nonstandard Hubbard terms as a function of the system parameters. Indeed, from Eq.~\eqref{Om1}, we notice that $\Omega_1$, as a function of the interaction range, becomes positive for $\bd\simeq L/2$, where $L$ is the well width. Indeed, the main contribution to the integral in Eq.~\eqref{Om1} comes from $x\simeq -L/2$, at the maximum of the left-orbital function. In this case, 
$$\Omega_1\propto \int\limits_{y_1}^{y_2} \Psi_L(y)\Psi_R(y)\,dy\,,$$
where $y_{1,2}=-L/2\mp\bd$. As a result, $\Omega_1\simeq 0$ for $\bd\simeq L/2$ due to orthogonality, see Eq.~\eqref{orth}. Subsequently, $\Omega_1$ starts to increase for $\bd\gtrsim L/2$, as the long-range interaction begins to connect the central regions of the two WFs.
\begin{figure}[t]
\centering
\includegraphics[width=8.6cm]{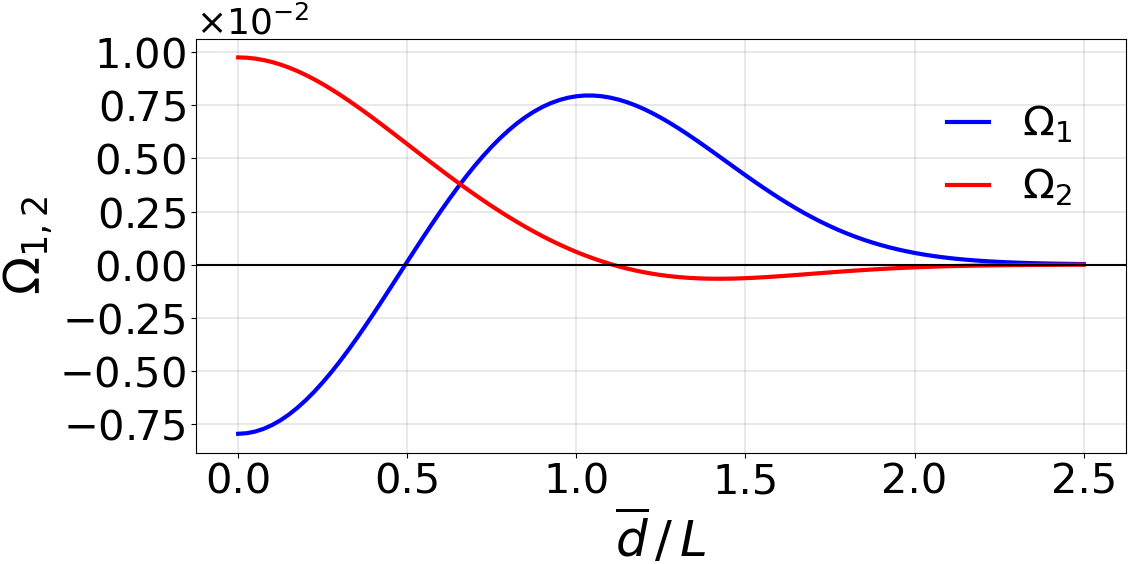}
\caption{DT amplitude $\Omega_1$ (blue curve) and PT amplitude $\Omega_2$ (red curve) as a function of the interaction range $\overline{d}$ for a symmetric square double-well potential. Parameters: $L=2$, $b=0.5$, $V_0=5$ and $\Omega_0=-0.22$. Interaction strength $V_\delta=1$, in arbitrary units.}
\label{fig4}
\end{figure}
This qualitative argument has been tested numerically in Fig.~\ref{fig4}, where 
the DT and PT terms ($\Omega_1$ and $\Omega_2$) for two distinguishable particles in a square double-well potential with long-range interaction are shown as a function of the rescaled interaction range $\bd/L$. 
For the sake of comparison with the contact interaction in Eq.~\eqref{delta}, in the calculations we kept $V_\delta=2\bd \,\bV$ fixed. It is clear that, in this way the results for the contact interaction are obtained in the limit $\bd\to 0$ and $\bV\to\infty$. The amplitudes are evaluated by substituting the exact WFs of Eq.~\eqref{Wannier_functionsb} in Eqs.~\eqref{Om1Om2}, by using the long-range potential in Eq.~\eqref{eq:long}. As expected, $\Omega_1$ undergoes a sign change for $\bd \gtrsim L/2$. Given that $\Omega_0<0$, see Eq.~\eqref{Wannier_functionsa}, the effective single-particle tunneling $\Omega_{eff}$ can be always suppressed for some finite interaction range $\bd\gtrsim L/2$ and a sufficiently large interaction strength $V_\delta$, since $\Omega_1 \propto V_\delta $.

In the next section, we will show how PT is still possible even in the case of single-particle tunneling suppression, due to a combined action of the nonstandard Hubbard DT term and the long-range interaction.

\subsection{Two interacting electrons with parallel spins in a square triple-well potential}
\label{subsec:NSH_B}
As we have discussed, the suppression of single-particle tunneling coupling in the nonstandard Hubbard model arises due to the interplay of long-range repulsive electron interaction and lattice potential. In principle, we would expect that this suppression, similarly to what happen in a flat band in twisted bilayer graphene systems \cite{bistritzer:moire_bands, vu:moire_mott, chan:pairing, cao:uncon_superc}, disrupt the electron transport. However, instead of being suppressed, transport can still occur via PT of localized electron pairs that are not subjected to ``decay'' through single-electron tunneling processes \cite{leggett:bookquantum}. 

One can argue that even in the context of the standard Hubbard model, single-electron hopping in a double-well potential is suppressed for large on-site interaction $(U)$. For this reason, it could be challenging to distinguish this suppression from the one due to the nonstandard DT term. To avoid this issue, let us consider two electrons with parallel spins so that they cannot occupy the same well due to the Pauli principle. In this case, the contribution of the long-range electron interaction in neighboring sites, $\bU\ll U$, replaces the standard on-site Hubbard term $U$. As a result, single-electron tunneling is not suppressed by the on-site interaction, while the DT term can still induce the suppression. Even in this scenario, similarly to the double-well case, the DT term and the single-particle tunneling term sum up to give an effective single-particle tunneling term $\Omega_{eff}$. Therefore, if $\Omega_0$ is exactly opposite to the DT term $\Omega_1$, the electron pair occupying two adjacent wells becomes stable and moves coherently due to the PT term.

\begin{figure*}[t]
\includegraphics[width=17.2cm]{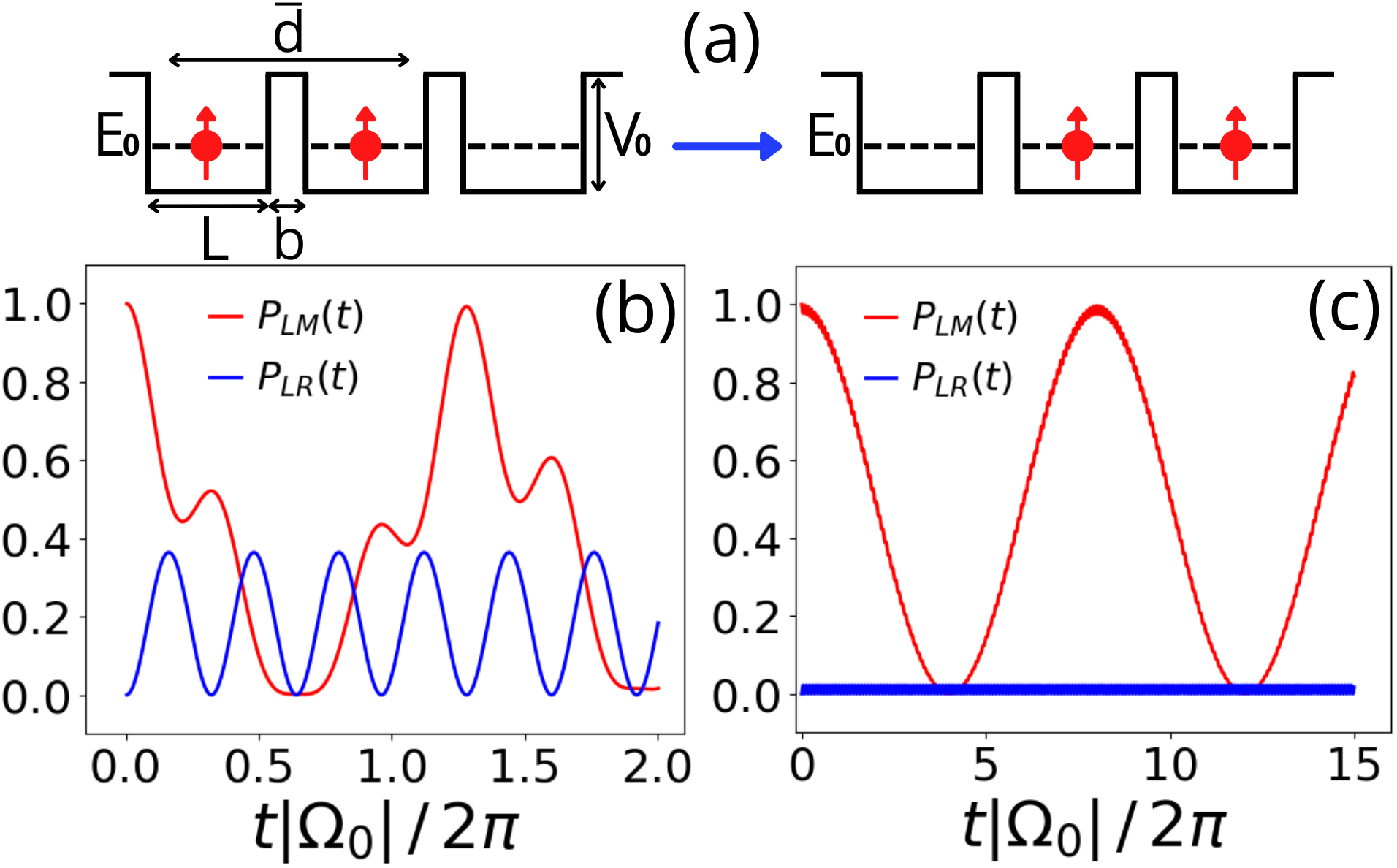}
\caption{(a) Coherent motion of two interacting electrons with parallel spins in a symmetric square triple-well potential, corresponding to the PT process. Black dashed lines correspond to the single-site ground state energy $E_0$, while $\bd$ is the interaction range. (b) Occupancy probabilities $P_{LM}(t)$ (red curve) and $P_{LR}(t)$ (blue curve) for $L=2$, $b=0.5$, $V_0=5$ and $\Omega_0\simeq-0.22$. (c) Occupancy probabilities $P_{LM}(t)$ (red curve) and $P_{LR}(t)$ (blue curve) for $L=4$, $b=1$, $V_0=5$ and $\Omega_0\simeq-0.0167$. Interaction strength $V_\delta=3$ and interaction range $\bd/L=2$, so that $\bU\simeq 0.35$, $\Omega_1\simeq 0.0125$ and $\Omega_2\simeq 0.0012$ in (b) and $\bU\simeq 0.19$, $\Omega_1\simeq 0.0027$ and $\Omega_2\simeq -6.4 \cdot 10^{-6}$ in (c), in arbitrary units.}
\label{fig5}
\end{figure*}

To show this mechanism explicitly, let us consider two electrons with parallel spins in a triple-well potential, as shown in Fig.~\ref{fig5} (a). The corresponding lowest-band Hamiltonian can be written as 
\begin{equation}
\hat{H}=\hat{H}_3+\hat{V}\,,
\label{eq:ns}
\end{equation}
where $\hat{H}_3$ is the non-interacting tight-binding tunneling Hamiltonian, given by Eq.~\eqref{tunn_ham3}, while $\hat{V}$ represents the inter-particle interaction term. The non-interacting Hamiltonian can be rewritten in the second quantization formalism as 
\begin{equation}
\hat H_3= E_0\sum_{j=1}^3\hn_j+\Omega_0\left(\ha_L^{\dagger}\ha_M+\ha_M^{\dagger}\ha_R+H.c.\right)\,,
\label{free_h0}
\end{equation}
where $\ha_j^{(\dagger)}$ destroys (creates) an electron at site $j=1,2,3\equiv L,M,R$, $\hn_j=\ha_j^\dagger\ha_j$ is the number operator, $E_0$ is the site-energy and $\Omega_0$ is the tunneling energy given by Eq.~\eqref{sqbard}. Since the Hamiltonian does not contain any spin-flip terms, the number operators $\hat{n}_j$ involve only parallel spins, so that the spin indices can be omitted.

In a similar way, the interaction operator $\hat{V}$ can be written in the second quantization formalism as
\begin{equation}
\hV=\frac{1}{2}\sum_{i'j'ij}V_{i'j'ij}\ha_{i'}^{\dagger}\ha_{j'}^{\dagger}\ha_j\ha_i\,,
\label{pot_electrons}
\end{equation}
where $V_{i'j'ij}$ is obtained by substituting in Eq.~\eqref{V_int_terms} the triple-well WFs $\Psi_j(x)\equiv\braket{x|\ha^\dagger_j|0}$ given by Eqs.~\eqref{wf3}, and the long-range potential interaction of Eq.~\eqref{eq:long}. Thus, considering only parallel-spin electron motion, Eq.~\eqref{pot_electrons} can be explicitly written as
\begin{equation}
\begin{aligned}
\hat{V}&=\bU\left(\hn_L\hn_M+\hn_M\hn_R\right)\\
&+\Omega_{1}\left(\hn_L\ha_M^{\dagger}\ha_R+\hn_R\ha_M^{\dagger}\ha_L+H.c.\right)\\
&-\Omega_2\left(\hn_M\ha_L^{\dagger}\ha_R+\hn_M\ha_R^{\dagger}\ha_L\right)\,,
\label{int4}
\end{aligned}
\end{equation}
where $\bU$ represents the nearest neighbor interaction term, obtained in the triple-well case by replacing $\Psi_R(x)$ with $\Psi_M(x)$ in Eq.~\eqref{uu}, so that
\begin{equation}
\bU=\int\Psi_L^2(x)V(x-y)\Psi_M^2(y)\,dx\,dy\,,
\label{uutriple}
\end{equation}
while the last two terms describe respectively the DT and PT processes, with amplitudes given by
\begin{subequations}
\label{3wint}
\begin{align}
&\Omega_1=\int \Psi_L^2(x)\Psi_M(y)V(x-y)\Psi_R(y)\,dx\,dy\,,
\label{3winta}\\
&\Omega_2=\int \Psi_L(x)\Psi_M(x)V(x-y)
\Psi_M(y)\Psi_R(y)\,dx\,dy\,,
\label{3wintb}
\end{align}
\end{subequations}
where $\Omega_1 \equiv \Omega_1^{M\to R}=\Omega_1^{M\to L}$ and $\Omega_2 \equiv \Omega_2^{L\to M, M\to R}=\Omega_2^{R\to M, M\to L}$. Notice that in our calculations, we have chosen the interaction range $\overline{d}$ in Eq.~\eqref{eq:long} so that the contribution from the next-to-nearest neighbor term can be neglected. In this way, the total Hamiltonian in Eq.~\eqref{eq:ns} represents the \textit{nonstandard} Hubbard model, whereas the \textit{extended} Hubbard model arises simply by setting $\Omega_1=\Omega_2=0$ in Eq.~\eqref{int4}. Finally, we observe that in the presence of long-range interaction, the DT term $\Omega_1$ changes its sign depending on the interaction range, as illustrated in Fig.~\ref{fig4} for the double-well system (for the triple-well case, see Fig.~\ref{fig_appendix_1} in the Supplemental Material \cite{supplmat}).
\begin{figure*}[t]
\includegraphics[width=17.2cm]{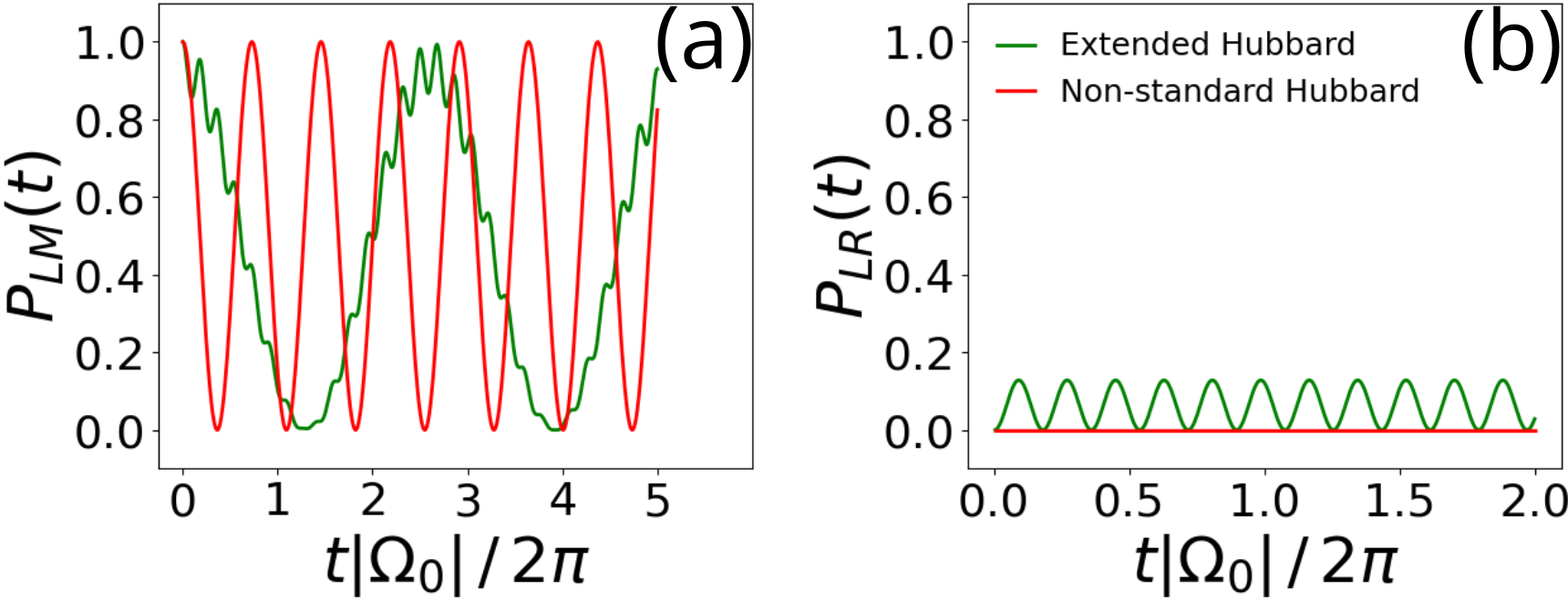}
\caption{(a) Occupancy probabilities $P_{LM}(t)$ and (b) occupancy probabilities $P_{LR}(t)$ obtained with the nonstandard Hubbard model (red curves) and the extended Hubbard model (green curves), for the case of complete single-particle tunneling suppression. Parameters: $L=2$, $b=0.1$, $V_0=1.1$ and $\Omega_0\simeq -0.32$. Interaction strength $V_\delta=22$ and interaction range $\bd/L=2$, so that $\bU\simeq 1.54$, $\Omega_1\simeq-\Omega_0$ and $\Omega_2 \simeq 0.22$, in arbitrary units.}
\label{fig6}
\end{figure*}

The effectiveness of our approach can be tested directly by studying the quantum dynamics of the system. In particular, let us consider as initial condition two electrons occupying two neighboring wells $\bj$ and $\bj'$. Their time-dependent wave function can always be written as
\begin{equation}
\ket{\Psi^{\left(\bj\,\bj'\right)}(t)}=\sum_{j<j'}b_{jj'}^{(\bj\,\bj')}(t)\,\ha_j^\dagger\ha_{j'}^\dagger \ket{0}\,,
\label{s0}
\end{equation}
where $j,j'=L,M,R,$ while the upper indices $(\bj\,\bj')$ label the initial state. Specifically, let us choose $\bj=L$ and $\bj'=M$, so that the left and middle wells are initially occupied. Then, Eq.~\eqref{s0} can be explicitly written as
\begin{equation}
\begin{aligned}
\ket{\Psi^{(LM)}(t)}=&\Big[b^{(LM)}_{LM}(t)\,\ha_L^\dagger\ha_M^\dagger + b^{(LM)}_{LR}(t)\,\ha_L^\dagger\ha_R^\dagger\\
&+b^{(LM)}_{MR}(t)\,\ha_M^\dagger\ha_R^\dagger\Big]\ket{0}\,.
\label{s1}
\end{aligned}
\end{equation}
By substituting Eq.~\eqref{s1} into the time-dependent Schr\"odinger equation
\begin{equation}
i\partial_t\ket{\Psi^{(\bj\,\bj')}(t)}=\left(\hat{H}_3+\hV\right)\ket{\Psi^{(\bj\,\bj')}(t)}\,,
\label{s1_5}
\end{equation}
we obtain the following equations of motion: 
\begin{equation}
\begin{aligned}
\label{3wells}
i\dot b_{LM}^{(LM)}(t)&=\left(2E_0+\bU\right)b_{LM}^{(LM)}(t)\\
&+\left(\Omega_0+\Omega_1\right)b_{LR}^{(LM)}(t)+\Omega_2 b_{MR}^{(LM)}(t)\,,\\
i\dot b_{LR}^{(LM)}(t)&=2E_0b_{LR}^{(LM)}(t)\\
&+\left(\Omega_0+\Omega_1\right)\left[b_{LM}^{(LM)}(t)+b_{MR}^{(LM)}(t)\right]\,,\\
i\dot b_{MR}^{(LM)}(t)&=\left(2E_0+\bU\right)b_{MR}^{(LM)}(t)\\
&+\left(\Omega_0+\Omega_1\right)b_{LR}^{(LM)}(t)+\Omega_2 b_{LM}^{(LM)}(t)\,.
\end{aligned}
\end{equation}
Looking at Eqs.~\eqref{3wells}, we notice that the DT term $\Omega_1$ appears only together with the single-particle tunneling $\Omega_0$, thus giving rise to an effective single-particle tunneling $\Omega_{eff}=\Omega_0+\Omega_1$ \cite{dutta:non_standard,jurgensen:density_induced_optical,jurgensen:observation_density_induced_tunnelling}. Eqs.~\eqref{3wells}
can be integrated numerically to obtain 
the occupancy probabilities for all sites of the triple-well system as a function of time. Specifically, the probability to find the two electrons in the wells $j,j'$ is defined as
\begin{equation}
\begin{aligned}
P_{jj'}(t)&=\bra{\Psi^{(LM)}(t)}\hn_j\hn_{j'}\ket{\Psi^{(LM)}(t)}\\
&=\left|b_{jj'}^{(LM)}(t)-b_{j'j}^{(LM)}(t)\right|^2\,, 
\label{s5}
\end{aligned}
\end{equation}
while the probability to find one electron occupying the well $j$ is defined as
\begin{equation}
P_{j}(t)=\sum_{j'\neq j}\left|b_{jj'}^{(LM)}(t)\right|^2\,.
\label{s6}
\end{equation}

In Fig.~\ref{fig5} (b,c), we show the probabilities $P_{LM}(t)$ and $P_{LR}(t)$, derived from Eq.~\eqref{s5}, for two different geometries of the triple-well system at some fixed inter-particle interaction strength. Particularly, in Fig.~\ref{fig5} (c), we adjust the geometry of the system (by enlarging the well and barrier widths) to produce a significant suppression of $P_{LR}(t)$, if compared with that in Fig.~\ref{fig5} (b). This suppression suggests the emergence of a propagating correlated electron pair within the system, showing that single-particle tunneling can be suppressed induced by modifying the well parameters. Note that a similar suppression is also observable within the extended Hubbard model framework. Specifically, it is easy to show that the suppression of $P_{LR}(t)$ in the extended Hubbard model occurs when $\bU \gg \Omega_0$ (for details, see the Supplemental Material \cite{supplmat}).

Clearly, the nonstandard and extended Hubbard model diverge significantly when complete suppression of single-particle tunneling occurs, i.e. for $\Omega_1=-\Omega_0$. To show this explicitly, we adjust the geometry of the system and the interaction strength to achieve complete suppression of single-particle tunneling $\Omega_{eff}$. Results are shown in Fig.~\ref{fig6} for both nonstandard (red curves) and extended (green curves) Hubbard models. As one can see, notable distinctions between the two models' predictions exist. Specifically, the extended Hubbard model predicts a small, but not zero, amplitude for $P_{LR}(t)$ (see Fig.~\ref{fig6} (b)), as well as a smaller oscillation frequency of $P_{LM}(t)$ compared to the nonstandard Hubbard model (see Fig.~\ref{fig6} (a)). Given that in the nonstandard Hubbard model the single particle tunneling is suppressed, the enhanced transport efficiency, signaled by the high frequency of oscillations of $P_{LM}(t)$ is due to the presence of the PT term.

Finally, one may wonder what is the region of parameters in which the nonstandard and extended Hubbard model give approximately similar outcomes. Within the validity of the single-band approximation, a glance at Eq.~\eqref{3wells} reveals that the two Hubbard models are expected to give close results when the $\Omega_1$ and $\Omega_2$ terms become negligible compared to $\Omega_0$, namely for sufficiently weak interaction strength. A detailed comparison between the two Hubbard models, as well as a comparison with our analytical approach, is reported in the Supplemental Material \cite{supplmat}.

\section{CONCLUSIONS}
In conclusion, we have explored the conditions governing the suppression of single-particle tunneling coupling in periodic systems, within the framework of a nonstandard Hubbard model, including density-induced tunneling and pair tunneling terms. Our findings demonstrate that such suppression cannot occur with a conventional contact repulsive interaction, but only in presence of a long-range repulsive interaction. A better understanding of the mechanism underlying the suppression of the single-particle tunneling could be a significant issue in the theory of quantum transport in correlated systems. Indeed, as we have shown here, see Fig.~\eqref{fig6}, in presence of single-particle tunneling suppression the dynamics is dominated by pair tunneling, which enhances the transport efficiency.

The consequences of these effects are far-reaching, since single-particle tunneling suppression and pair tunneling dominated dynamics may lead to novel transport regimes, characterized by efficient and robust electron pair transport. Indeed, within the nonstandard Hubbard model considered here, increasing the interaction strength not only suppresses single-particle tunneling but also enhances pair tunneling, introducing a competition between these two effects. Such interplay may lead to nontrivial transport regimes that could potentially expand the paradigm of Mott-insulator transitions \cite{zhou:pair_tunnelling_bosons} beyond the standard Hubbard model. In the future, we plan to investigate the impact of the effects unveiled in this manuscript in lattice models of different dimensions.

\begin{acknowledgments}
FB, MZ and GLC acknowledge the support of the Iniziativa Specifica INFN-DynSysMath. This work has been financially supported by the Catholic University of Sacred Heart and by M.I.U.R. within the Project No. PRIN 20172H2SC4. MZ acknowledges the Ermenegildo Zegna's Group for the financial support. SG would like to thank Yuval Oreg and Erez Berg for the helpful discussions and suggestions. We also thank Samy Mailoud and Guido Farinacci for the discussions and for their valuable contribution at the initial stage of this work.
\end{acknowledgments}


%

\onecolumngrid
\newpage

\setcounter{page}{1}
\setcounter{section}{0}

\begin{center}
\textbf{Supplemental Material for ``Nonstandard Hubbard model and electron pairing''}
\end{center}

\section{Two-potential approach to the bound-state spectrum and the Bardeen formula}
\label{appendix_A}
\renewcommand{\theequation}{S1.\arabic{equation}}
\setcounter{equation}{0}
\renewcommand{\thefigure}
{S1.\arabic{figure}}
\setcounter{figure}{0}

The standard perturbation approach consists in separating the total Hamiltonian $\cH$ of an entire system into an ``unperturbed'' Hamiltonian and a ``perturbation''. Considering the symmetric double-well potential in Fig.~\ref{fig1} (a) of the main text, we choose $\cH= H_1+\cV_2$, where $H_1=\cK+\cV_1$ is the left-well Hamiltonian and $\cV_2(x)=\cV_1(-x)$ is the right-well potential ($\cK = -\nabla^2/2m $ is the kinetic part). Such a separation implies that the spectrum of the unperturbed left-well Hamiltonian,
\begin{equation}
H_1\ket{\Phi_n^{(1)}}=E_n^{(1)}\ket{\Phi_n^{(1)}}\,,
\end{equation}
must be known (here $E_n^{(1)}$ denotes discrete and continuum spectrum states). The total Hamiltonian spectrum $\cH\ket{\psi_n}=E_n\ket{\psi_n}$ can be obtained by solving the following equations \cite{thouless:book}
\begin{subequations}
\label{p1}
\begin{align}
&E_n=E_n^{(1)}+\braket{\Phi_n^{(1)}|\cV_2|\Phi_n^{(1)}}
+\braket{\Phi_n^{(1)}|\cV_2\,\tG(E_n)\,\cV_2|\Phi_n^{(1)}}\,,
\label{p1a}\\
&\ket{\psi_n}=\ket{\Phi_n^{(1)}}+\tG(E_n)\,
\cV_2\ket{\Phi_n^{(1)}}\,,
\label{p1b}
\end{align}
\end{subequations}
where $$\tG(E_n)=\left(1-\Lambda_n^{(1)}\right)(E_n-\cH)^{-1}$$ represents the total Green's function and $\Lambda_n^{(1)}=\ket{\Phi_n^{(1)}}\bra{\Phi_n^{(1)}}$ is a projection operator on the state $E_n$ of the Hamiltonian $H_1$. For simplicity, in the following we will consider the ground state $(n=0)$ of the potential $\cV_1$. 

The Green's function $\tG(E)$ can be obtained directly from the Lippmann-Schwinger equation
\begin{equation}
\tG(E)=\left[1+\tG(E)\,\cV_2\right]\tG_1(E)\,, 
\label{pert1}
\end{equation}
where
\begin{equation}
\tG_1(E)=\left(1-\Lambda_0^{(1)}\right){\frac{1}{E-H_1}}=\sum_{n\not =0}\frac{\ket{\Phi_{n}^{(1)}}\bra{\Phi_{n}^{(1)}}}{E-E_{n}^{(1)}}\,.
\label{pert1b}
\end{equation}

By solving Eq.~\eqref{p1a} with respect to $E_n$, we find the energy spectrum of the system, while the corresponding eigenstates $\ket{\psi_n}$ are obtained from Eq.~\eqref{p1b}. For the treatment of Eqs.~\eqref{p1}, we use the perturbative expansion obtained by iterating the Lippmann-Schwinger equation in Eq.~\eqref{pert1}, namely
\begin{equation}
\tG(E)=\tG_1+\tG_1\,\cV_2\,\tG_1+\tG_1\,\cV_2\,\tG_1\,\cV_2\,\tG_1+\dots\,.
\label{sp1}
\end{equation}
Substituting Eq.~\eqref{sp1} into Eqs.~\eqref{p1}, we find the Brillouin-Wigner perturbation series \cite{thouless:book} for the energy spectrum of the Hamiltonian $\cH$ in powers of the perturbation $\cV_2$. In particular, Eq.~\eqref{p1a} at the second order in $\cV_2$, is given by
\begin{equation}
\begin{aligned}
E&=E_0^{(1)}+\braket{\Phi_0^{(1)}|\cV_2|\Phi_0^{(1)}}\\
&+\braket{\Phi_0^{(1)}|\sum_{n\not =0}\cV_2\frac{\ket{\Phi_{n}^{(1)}}\bra{\Phi_{n}^{(1)}}}{E-E_{n}^{(1)}}\cV_2|
\Phi_0^{(1)}} +\cO \left(\cV_2^3\right)\,.
\label{sp2}
\end{aligned}
\end{equation}

Looking at Eqs.~\eqref{p1} and \eqref{sp1}, we notice that the main problem with the perturbative treatment is the absence of a small parameter in the corresponding expansions. At first sight, we could consider the second order perturbation term in $\cV_2$, see Eq.~\eqref{sp2}, as a small parameter. This term is suppressed because the wave function $\Phi_0^{(1)}(x)$ decreases exponentially for $x>0$, where $\cV_2(x)$ is large, see Fig.~\ref{fig1} (c) of the main text. However, the higher order terms of the expansion include an overlap of the potential $\cV_2(x)$ with the wave functions $\Phi_n^{(1)}(x)$ of the continuum spectrum, which are not suppressed at all for large $|x|$. This makes the expansion in Eq.~\eqref{sp2} not applicable for evaluating the eigenspectrum of the double-well potential. In general, this is not surprising, since any problem related to tunneling is usually a non-perturbative one.

However, we can use a different treatment of the Green's function $\tG(E)$, which leads to a perturbative series in powers of an effectively small expansion parameter. Such a two potential approach was originally developed for tunneling to the continuum in \cite{gurvitz:novel_approach,gurvitz:decay_width, gurvitz:modified_pot_approach, gurvitz:twoelectroncorrelated}, and we have extended it to bound-state problems.

Consider the total Green's function $G(E)$ of the double-well system in Fig.~\ref{fig1} (a) of the main text. It contains two poles for $E=\cE_{1,2}$, corresponding to the two eigenstates of the system with energies close to $E_0$. Comparing $G(E)$ with $\tG(E)=\left(1-\Lambda_0^{(1)}\right)G(E)$, we observe that the two Green's functions are indeed very similar. The only difference is related to the projection operator $\left(1-\Lambda_0^{(1)}\right)$, which excludes the ground state $\ket{\Phi_0^{(1)}}$ from the spectral representation. However, the ground state $\ket{\Phi_0^{(2)}}$ of the right well is not excluded by the projection operator. This state would dominate the Green's function behavior at $E\simeq E_0$, making it close to the Green's function $G_2(E)=(E-H_2)^{-1}$ of the second well, which is given by
\begin{equation}
G_2(E)=\frac{\ket{\Phi_0^{(2)}}\bra{\Phi_0^{(2)}}}{E-E_0}+\sum_{n\not=0}\frac{\ket{\Phi_n^{(2)}}\bra{\Phi_n^{(2)}}}{E-E_n^{(2)}}\,.
\label{g}
\end{equation}
This suggests a new expansion of the Green's function $\tG(E)$ in terms of $G_2(E)$. To find it, we multiply the Lippmann-Schwinger equation~\eqref{pert1} by $(E-H_{1})$, thus obtaining 
$$\tG(E)(E-H_1)=\left[1+\tG (E)\,\cV_2\right]\left(1-\Lambda_0^{(1)}\right)\,.$$
Using $E-H_{1}\equiv E-H_{2}+\cV_2-\cV_1$, we can write
\begin{equation}
\begin{aligned}
\tG(E)(E-H_2)&=\tG(E)\,(\cV_1-\cV_2)\\
&+\left[1+\tG(E)\,\cV_2\right]\left(1-\Lambda_0^{(1)}\right)\,.
\label{0}
\end{aligned}
\end{equation}
Multiplying Eq.~\eqref{0} by $G_2(E)$, we obtain
\begin{equation}
\begin{aligned}
\tG(E)=&\left[1+\tG(E)\,\cV_1\right]G_2(E)\\
&-\left[1+\tG(E)\, \cV_2\right]\Lambda_0^{(1)}G_2(E)\,.
\label{1}
\end{aligned}
\end{equation}
Eq.~\eqref{1} shows the exact relation between the Green's function $\tG(E)$ and the Green's function of the second well $G_2(E)$ of Eq.~\eqref{g}. Notice that, in the limit $E\to E_0$,
\begin{equation}
\Lambda_0^{(1)}G_2(E)\stackrel{E\to E_0}{\Longrightarrow}\beta \,\frac{\ket{\Phi_0^{(1)}}\bra{\Phi_0^{(2)}}}{E-E_0}\,,
\label{ov}
\end{equation}
where $\beta = \braket{\Phi_0^{(1)}|\Phi_0^{(2)}}$ represents the overlap of the two (non-orthogonal) wave functions of neighboring sites. Since the site wave functions are mainly localized in the respective wells, their overlap $\beta\ll 1$, as can be explicitly shown in the semi-classical limit. Indeed, the left-well orbital function can be written as
\begin{equation}
\Phi_0^{(1)}(x)=
\begin{cases}
\Phi_0^{(1)}(-\bx)\,e^{-\int\limits_{-\bx}^x p(x')\,dx'} &{\rm for} -\bx < x\le 0\\\
\Phi_0^{(1)}(0)\,e^{-p(0)x} &{\rm for}\quad x> 0
\end{cases}\,,
\label{wf1}
\end{equation}
where $p(x)=\sqrt{2m\left(\cV_1(x)-E_0\right)}$ is the (imaginary) momentum under the barrier, so that $p(0)=\sqrt{-2mE_0}$, and $-\overline{x}$ is the classical turning point, with $\cV_{1,2}(\mp \,{\overline{x}})=E_0$, see Fig.~\ref{fig1} (b,c) of the main text. We obtain the same expression for the right-well orbital function $\Phi_0^{(2)}(x)$, under the substitution $x\to -x$ and $\bx\to -\bx$. Therefore, we obtain
\begin{equation}
\begin{aligned}
\beta =&2\int\limits_{-\infty}^0 \Phi_0^{(1)}(x)\Phi_0^{(2)}(x)\,dx\simeq \left[\Phi_0^{(1)}(-\bx)\right]^2\\
&\times e^{-\int\limits_0^{\bx}p(x')\,dx'} \int\limits_{-\bx}^0e^{-\int\limits_{-\bx}^x p(x')\,dx'+p(0)x}\,dx\,,
\label{beta}
\end{aligned}
\end{equation}
where we have neglected the integration region $(-\infty,-\bx$), whose contribution is exponentially small. To perform the integration in Eq.~\eqref{beta}, we apply the stationary phase approximation. The stationary point of the variable $x$ is obtained by differentiating the exponential factor, so that 
$$-p(x)+p(0)=0.$$
Solving this equation, we find that the stationary point corresponds to $x=0$. As a result, the integral over $x$ in Eq.~\eqref{beta} is given by $C\exp\bigg(-\int\limits_{-\bx}^0 p(x')\,dx'\bigg)$, where $C$ is a pre-exponential factor of the stationary phase approximation. Finally, we obtain from Eq.~\eqref{beta}:
\begin{equation}
\beta\simeq 2C \left[\Phi_0^{(1)}(-\bx)\right]^2e^{-\int\limits_{-\bx}^{\bx} p(x')\,dx'}\propto T_0\,,
\label{beta1}
\end{equation}
where $T_0$ is the barrier penetration coefficient in Eq.~\eqref{T0_barr_coeff} of the main text. Considering Eq.~\eqref{1} in the limit $E\to E_0$, and using Eq.~\eqref{g}, $\tG(E)$ can be written as
\begin{equation}
\begin{aligned}
\tG(E) &=\left(1+\tG(E) \,\cV_1\right)\,\frac{\ket{\Phi_0^{(2)}}\bra{\Phi_0^{(2)}}}{E-E_0}\\
&-\beta \left(1+\tG(E) \cV_2\right)\frac{\ket{\Phi_0^{(1)}}\bra{\Phi_0^{(2)}}}{E-E_0}\,.
\label{2}
\end{aligned}
\end{equation}
Eq.~\eqref{2} can be easily solved for the zero-order term in $\beta$, obtaining 
\begin{equation}
\tG(E)\simeq\frac{\ket{\Phi_0^{(2)}}\bra{\Phi_0^{(2)}}}{E-E_0-\bom}+\cO \left(\beta \right)\,,
\label{9}
\end{equation}
where
\begin{equation}
\begin{aligned}
\bom &=\braket{\Phi_0^{(1)}|\cV_2|\Phi_0^{(1)}}
=\braket{\Phi_0^{(2)}|\cV_1|\Phi_0^{(2)}}\\
&=\int\limits_{-\infty}^\infty
\left[\Phi_0^{(1)}(x)\right]^2\,\cV_2(x)\,dx
\label{bom}
\end{aligned}
\end{equation}
is the diagonal energy shift. Since the potential $\cV_2(x)$ overlaps with the orbital function tail, see Eq.~\eqref{wf1}, $\bom\propto T_0^2\propto\beta^2$ (c.f. with Eqs.~\eqref{beta} and \eqref{beta1}) and therefore it can be neglected. 

Substituting Eq.~\eqref{9} into Eq.~\eqref{p1a}, we find that the eigenstate energies of the system $E_\pm$ (up to the $\cO\left(\beta^2\right)$ terms) are obtained from the equation $$E-E_0=\Omega_0^2/(E-E_0)\,,$$ giving $E_\pm=E_0\pm\Omega_0$, where 
\begin{equation}
\Omega_{0}=\bra{\Phi_0^{(1)}}\cV_2\ket{\Phi_0^{(2)}}=
\int\limits_{0}^\infty\Phi_0^{(1)}(x)\,\cV_2(x)\,\Phi_0^{(2)}(x)\,dx
\label{11}
\end{equation}
is the off-diagonal energy shift, corresponding to the energy split between the two lowest eigenstates, $\Omega_0=(E_+-E_-)/2$, and represents the tunneling coupling energy. Notice that $\Omega_0<0$, since $\cV_2(x)<0$, see Fig.~\ref{fig1} of the main text. Using the Schr\"odinger equation $$\cV_2\ket{\Phi_0^{(2)}}=(E_0-\cK)\ket{\Phi_0^{(2)}}\,,$$ and $$\Phi_0^{(1)}(x)=\Phi_0^{(1)}(0)e^{-\sqrt{-2mE_0}x}$$ for $x\ge 0$, we can evaluate the integral in Eq.~\eqref{11} by integrating by parts, obtaining 
\begin{equation}
\begin{aligned}
\Omega_{0} &=
\int\limits_{0}^\infty \Phi_0^{(1)}(x)\left (E_0+\frac{1}{2m}\frac{d^2}{dx^2}\right ) \Phi_0^{(2)}(x)\\
&=\frac{1}{2m}\left[\Phi_0^{(1)\prime}(0)\Phi_0^{(2)}(0)-\Phi_0^{(1)}(0)\Phi_0^{(2)\prime}(0)\right]\,,
\end{aligned}
\label{12}
\end{equation}
where $\Phi_0^{(1,2)\prime}(0)=(d/dx)\Phi_0^{(1,2)}(x)\big|_{x\to 0}$. This equation represents the Bardeen formula \cite{bardeen:tunnelling_many_body}, although we use different orbital potentials, namely $\cV_{1,2}(x)=0$ beyond the separation point, see Fig.~\ref{fig1} (b,c) of the main text (c.f. \cite{gurvitz:novel_approach,gurvitz:decay_width, gurvitz:modified_pot_approach}). The latter gives us $$\Phi_0^{(1,2)}(x)=\Phi_0^{(1,2)}(0)\,e^{\mp\sqrt{-2mE_0}x}$$ for $x\gtrless 0$, so that $$\Phi_0^{(1,2)\prime}(0)=\mp\sqrt{-2mE_0}\Phi_0^{(1,2)}(0)\,.$$
Substituting this result into Eq.~\eqref{12}, we obtain the following simple expression for the tunneling energy:
\begin{equation}
\Omega_0=
-\sqrt{\frac{2|E_0|}{m}}\Phi_0^{(1)}(0)\Phi_0^{(2)}(0)\,.
\label{13}
\end{equation}
Evaluating Eq.~\eqref{13} in the semiclassical limit, using Eq.~\eqref{wf1} for the orbital functions, we obtain (c.f. with Eq.~\eqref{beta1})
\begin{equation}
\Omega_0\simeq -\sqrt{\frac{2|E_0|}{m}}\left[\Phi_0^{(1)}(-\bx)\right]^2e^{-\int\limits_{-\bx}^{\bx} p(x')\,dx'}\propto \beta\propto T_0\,.
\label{14}
\end{equation}
Therefore, similarly to the overlap integral $\beta$, the tunneling coupling $\Omega_0$ is also proportional to the penetration coefficient $T_0$. Notice that Eq.~\eqref{14} for the tunneling energy $\Omega_0$ was obtained by keeping the first (zero order) term in the expansion of $\tG(E)$ in powers of $\beta$. The accuracy of Eqs.~\eqref{12} and \eqref{13} is therefore up to the terms $\cO \left(\beta^2\right)$.

\section{Exact equations of motion of two electrons in a symmetric triple-well potential}
\label{appendix_B}
\renewcommand{\theequation}{S2.\arabic{equation}}
\setcounter{equation}{0}
\renewcommand{\thefigure}
{S2.\arabic{figure}}
\setcounter{figure}{0}
\begin{figure*}[t]
\includegraphics[width=17.2cm]{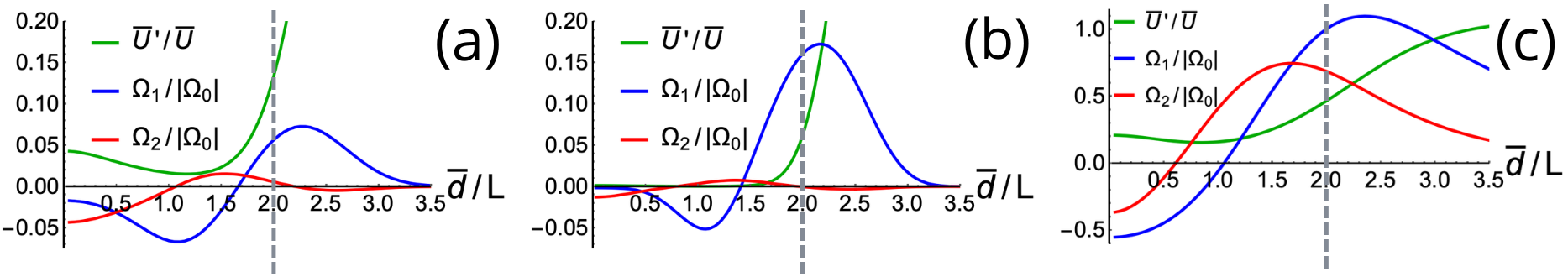}
\caption{Next-to-nearest neighbor interaction normalized by the nearest neighbor interaction $\bU'/\bU$ (green curves), DT amplitude normalized by the single-particle tunneling $\Omega_1/|\Omega_0|$ (blue curves) and PT amplitude normalized by the single-particle tunneling $\Omega_2/|\Omega_0|$ (red curves) as a function of the interaction range $\overline{d}$, for the three different geometries of the square triple-well potential presented in Fig.~\ref{fig5} (b,c) and Fig.~\ref{fig6} of the main text, respectively. (a) Parameters: $L=2$, $b=0.5$, $V_0=5$ and $\Omega_0\simeq-0.22$. Interaction strength $V_\delta=3$. (b) Parameters: $L=4$, $b=1$, $V_0=5$ and $\Omega_0\simeq-0.0167$. Interaction strength $V_\delta=3$. (c) Parameters: $L=2$, $b=0.1$, $V_0=1.1$ and $\Omega_0\simeq-0.32$. Interaction strength $V_\delta=22$, in arbitrary units. Vertical dashed lines represent the interaction range $\bd/L=2$, common to all panels.}
\label{fig_appendix_1}
\end{figure*}
Consider the triple-well potential $\cV(x)$ depicted in Fig.~\ref{fig2} (b) of the main text, where the lowest band is composed of three eigenstates $\psi_k(x)$, indexed by $k=1,2,3$ as described in Eq.~\eqref{eq:ham}. These eigenstates correspond to energies $\cE_k$, with $\cE_1<\cE_2<\cE_3<0$. We derive the tunneling Hamiltonian $\hat{H}_3$ for this band, expressed in the basis of Wannier functions $\ket{\Psi_j}$, with $j=1,2,3$ denoting the left, middle, and right wells, respectively. Accounting for the symmetry of $\cV(x)$, the tunneling Hamiltonian is given by
\begin{equation}
\begin{aligned}
\hat{H}_3 &=\bE_0\left(\hat{a}_L^\dagger \hat{a}_L+\hat{a}_M^\dagger \hat{a}_M+\hat{a}_R^\dagger \hat{a}_R\right)\\
&+\bom_0\left(\hat{a}_L^\dagger \hat{a}_M +\hat{a}_M^\dagger \hat{a}_R+H.c.\right)\\
&+\bom'_0\left(\hat{a}_L^\dagger \hat{a}_R+H.c.\right)\,,
\label{eq:FH}
\end{aligned}
\end{equation}
where $\bE_0$ represents the single-site energy, $\bom_0$ denotes nearest neighbor hopping, and $\bom'_0$ accounts for next-to-nearest neighbor hopping (which is usually disregarded, since it is a higher-order term in the penetration coefficient). In particular, Eq.~\eqref{eq:FH} constitutes a more general form than the nearest-neighbor Hamiltonian discussed in Eq.~\eqref{tunn_ham3} of the main text. The unitary matrix $R$, defined up to $\cO\left(\bom'_0/\bom_0\right)^2$ terms as
\begin{equation}
\label{untrexpl}
R=
\begin{pmatrix}
&\frac{1}{2} & \left(\frac{1}{\sqrt{2}}-\frac{\bom'_0}{4\bom_0}\right) & \frac{1}{2}\\
&\frac{1}{\sqrt{2}} & 0 & -\frac{1}{\sqrt{2}}\\
&\frac{1}{2} & -\left(\frac{1}{\sqrt{2}}+\frac{\bom'_0}{4\bom_0}\right) & \frac{1}{2}
\end{pmatrix}\,,
\end{equation}
diagonalizes $\hat{H}_3$, yielding eigenstates that can be identified with the exact solutions of the Schr\"odinger equation described in Eq.~\eqref{eq:ham}. Expressing the energies $\cE_k$ in terms of $\bE_0$, $\bom_0$ and $\bom'_0$, we get
\begin{equation}
\label{exact_sol_app}
\begin{aligned}
&\cE_{1,3}=\bE_0\mp\sqrt{2\bom_0^2+\left(\frac{\bom'_0}{2}\right)^2}+\frac{\bom'_0}{2}\,,\\
&\cE_2=\bE_0-\bom'_0\,.
\end{aligned}
\end{equation}
From Eqs.~\eqref{exact_sol_app}, up to $\cO\left(\bom'_0/\bom_0\right)^2$ terms, we get
\begin{equation}
\begin{aligned}
&\bE_0=\frac{\cE_1+\cE_2+\cE_3}{3}\,,\\
&\bom_0=\frac{\cE_1-\cE_3}{2\sqrt{2}}\,,\\
&\bom'_0=\frac{\cE_1-2\cE_2+\cE_3}{3}\,.
\label{sp3}
\end{aligned}
\end{equation}
Furthermore, the corresponding WFs $\Psi_j(x)$, expressed in terms of the lowest-band eigenstates $\psi_k(x)$ through the unitary transformation detailed in Eq.~\eqref{untrexpl}, are explicitly given by
\begin{equation}
\label{sp4}
\begin{aligned}
&\Psi_L(x)=\frac{1}{2}\psi_1(x)+\left(\frac{1}{\sqrt{2}}-\frac{\bom'_0}{4\bom_0}\right)\psi_2(x)
+\frac{1}{2}\psi_3(x)\,,\\
&\Psi_M(x)=\frac{1}{\sqrt{2}}\Big[\psi_1(x)
-\psi_3(x)\Big]\,,\\
&\Psi_R(x)=\frac{1}{2}\psi_1(x)-\left(\frac{1}{\sqrt{2}}+\frac{\bom'_0}{4\bom_0}\right)\psi_2(x)
+\frac{1}{2}\psi_3(x)\,.
\end{aligned}
\end{equation}
By neglecting $\cO \left(\bom'_0/\bom_0\right)$ terms, Eqs.~\eqref{sp3} and \eqref{sp4} coincide with Eqs.~\eqref{wf3} of the main text.

Moving on, the two-electron interacting term $\hat V$, described in Eq.~\eqref{V_int_terms} of the main text, is expressed in second quantization formalism as
\begin{equation}
\begin{aligned}
\hat{V}&=\bU\left(\hn_L\hn_M+\hn_M\hn_R\right)+\bU'\hn_L\hn_R\\
&+\Omega_{1}\left(\hn_L\ha_M^{\dagger}\ha_R+\hn_R\ha_M^{\dagger}\ha_L+H.c.\right)\\
&-\Omega_2\left(\hn_M\ha_L^{\dagger}\ha_R+H.c.\right)\,,
\label{int5}
\end{aligned}
\end{equation}
where $\bU$ denotes interaction between nearest neighbor sites, see Eq.~\eqref{uutriple} of the main text, $\bU'$ represents interaction between next-to-nearest neighbor sites, defined as
\begin{equation}
\bU'=\int\Psi_L^2(x)V(x-y)\Psi_R^2(y)\,dx\,dy\,,
\end{equation}
and $\Omega_1$ and $\Omega_2$ characterize the DT and PT processes, respectively, see Eqs.~\eqref{3wint} of the main text. 

Now, considering two electrons with parallel spins in the triple-well system, the total time-dependent wave function $\ket{\Psi^{(\bj\bj')}(t)}$ can be written as
\begin{equation}
\ket{\Psi}=\left[b_{LM}\ha_L^\dagger\ha_M^\dagger+b_{LR}\ha_L^\dagger\ha_R^\dagger+b_{MR}\ha_M^\dagger\ha_R^\dagger\right]\ket{0}\,,
\label{s1p}
\end{equation}
omitting both the upper indices $(LM)$, which denote the initial occupation of the system, and the time dependency. Substituting Eq.~\eqref{s1p} into the time-dependent Schr\"odinger Eq.~\eqref{s1_5} of the main text, considering $\hat H_3$ and $\hat V$ as described in Eqs.~\eqref{eq:FH} and \eqref{int5}, and applying the anti-commutation relations for the Fermi operators
$$\{\ha_j^\dagger \ha_{j'}\}=\delta_{jj'}\,,\quad\{\ha_j^\dagger \ha_{j'}^\dagger\}
=\{\ha_j \ha_{j'}\}=0\,,$$
we derive
\begin{equation}
\begin{aligned}
&\hH_3\ket{\Psi}=\hH_3\left[b_{LM}\ha_L^\dagger\ha_M^\dagger+b_{LR}\ha_L^\dagger\ha_R^\dagger+b_{MR}\ha_M^\dagger\ha_R^\dagger\right]\ket{0}\\
&=2\bE_0\left[b_{LM}\ha_L^\dagger\ha_M^\dagger+b_{LR}\ha_L^\dagger\ha_R^\dagger+b_{MR}\ha_M^\dagger\ha_R^\dagger\right]\ket{0}\\
&+\bom_0\left[b_{LM}\ha_L^\dagger\ha_R^\dagger+b_{LR}\left(\ha_L^\dagger\ha_M^\dagger+\ha_M^\dagger\ha_R^\dagger\right)+b_{MR}\ha_L^\dagger\ha_R^\dagger\right]\ket{0}\\
&-\bom'_0\left[b_{LM}\ha_M^\dagger\ha_R^\dagger+b_{MR}\ha_L^\dagger\ha_M^\dagger\right]\ket{0}\,,
\end{aligned}
\end{equation}
as well as
\begin{equation}
\begin{aligned}
&\hV\ket{\Psi}=\hV\left[b_{LM}\ha_L^\dagger\ha_M^\dagger+b_{LR}\ha_L^\dagger\ha_R^\dagger+b_{MR}\ha_M^\dagger\ha_R^\dagger\right]\ket{0}\\
&=\bU\left[b_{LM}\ha_L^\dagger\ha_M^\dagger+b^{(LR)}_{MR}\ha_M^\dagger\ha_R^\dagger\right]\ket{0}+\bU'\left[b_{LR}\ha_L^\dagger\ha_R^\dagger\right]\ket{0}\\
&+\Omega_1\Big[b_{LM}\ha_L^\dagger\ha_R^\dagger+b_{LR}\left(\ha_L^\dagger\ha_M^\dagger+\ha_M^\dagger\ha_R^\dagger\right)\\
&+b_{MR}\ha_L^\dagger\ha_R^\dagger\Big]\ket{0}
+\Omega_2\left[b_{LM}\ha_M^\dagger\ha_R^\dagger+b_{MR}\ha_L^\dagger\ha_M^\dagger\right]\ket{0}\,.
\end{aligned}
\end{equation}
Therefore, the resulting equations of motion are:
\begin{equation}
\label{3wells_complete}
\begin{aligned}
i\dot b_{LM}(t)&=\left(2E_0+\bU\right)b_{LM}(t)\\
&+\left(\bom_0+\Omega_1\right)b_{LR}(t)+\left(-\bom_0'+\Omega_2\right) b_{MR}(t)\,,\\
i\dot b_{LR}(t)&=\left(2E_0+\bU'\right)b_{LR}(t)\\
&+\left(\bom_0+\Omega_1\right)\left[b_{LM}(t)+b_{MR}(t)\right]\,,\\
i\dot b_{MR}(t)&=\left(2E_0+\bU\right)b_{MR}(t)\\
&+\left(\bom_0+\Omega_1\right)b_{LR}(t)+\left(-\bom_0'+\Omega_2\right)b_{LM}(t)\,.
\end{aligned}
\end{equation}
Notably, neglecting $\cO \left(\bom'_0/\bom_0\right)$ and $\cO \left(\bU'/\bU\right)$ terms, which are respectively next-to-nearest neighbor contributions to free and interacting dynamics, Eqs.~\eqref{3wells_complete} coincide with Eqs.~\eqref{3wells} of the main text. Finally, we investigate the effect of the interaction range $\overline{d}$ on the three terms $\bU'$, $\Omega_1$, and $\Omega_2$ in Fig.~\ref{fig_appendix_1}. Here, we keep the same geometries and interactions of Fig.~\ref{fig5} (b,c) and Fig.~\ref{fig6} of the main text, respectively. It is evident that in Fig.~\ref{fig_appendix_1} (a,b) the term $\bU'/U$ is negligible, while in Fig.~\ref{fig_appendix_1} (c) it starts to be relevant for the chosen interaction range. Despite that, it does not qualitatively affects the dynamics of the probabilities, confirming the validity of our approximation for the parameters considered in Fig.~\eqref{fig6} of the main text. Finally, the behavior of the DT and PT amplitudes, denoted as $\Omega_1$ and $\Omega_2$ respectively, mirrors that observed in the double-well scenario illustrated in Fig.~\ref{fig4} of the main text. Specifically, the DT term becomes positive for a sufficiently large interaction range, such that the chosen interaction strength makes $\Omega_1\simeq-\Omega_0$, confirming our hypothesis of single-particle tunneling suppression.

\section{Comparison between the extended and the nonstandard Hubbard model}
\label{appendix_C}
\renewcommand{\theequation}{S3.\arabic{equation}}
\setcounter{equation}{0}
\renewcommand{\thefigure}
{S3.\arabic{figure}}
\setcounter{figure}{0}
Let us compare the coherent dynamics of two interacting electrons with parallel spins within the symmetric square triple-well potential $\cV(x)$ shown in Fig.~\ref{fig2} (b) of the main text, using both the extended and the nonstandard Hubbard model.

For the extended Hubbard model, an analytical expression for the probability $P_{LR}(t)$ can be easily derived. Indeed, from Eqs.~\eqref{3wells} and \eqref{s5} of the main text, setting $\Omega_1=\Omega_2=0$, we obtain:
\begin{equation}
\label{app:anP}
P_{LR}(t)=\frac{4 \Omega_0^2}{8 \Omega_0^2 + \bU^2}\sin^2\left[\frac{\sqrt{8 \Omega_0^2 + \bU^2}}{2}t\right]\,.
\end{equation}

\begin{figure*}[t]
\includegraphics[width=17.2cm]{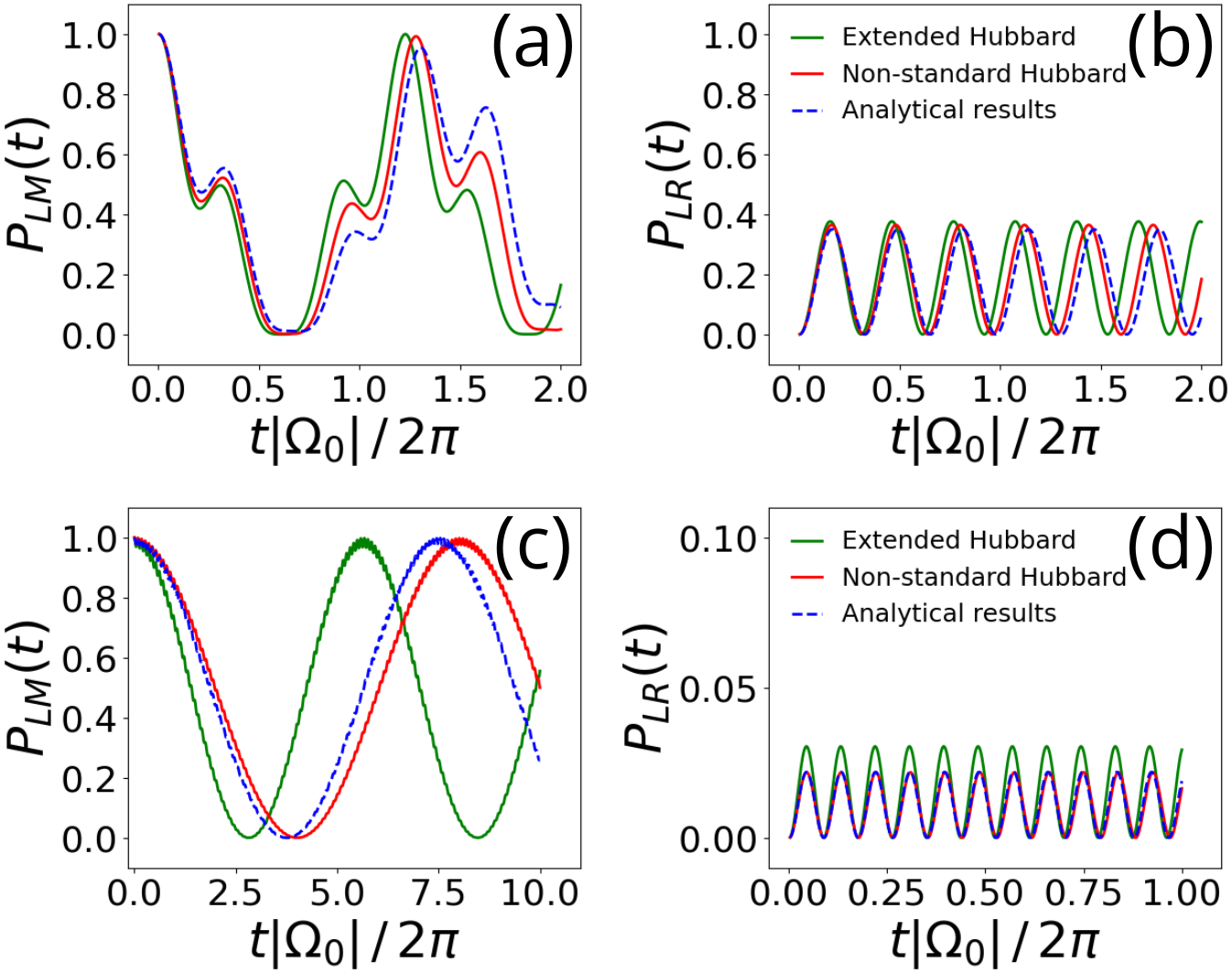}
\caption{Occupancy probabilities $P_{LM}(t)$ (left panels) and $P_{LR}(t)$ (right panels) for two electrons with parallel spins in a triple-well potential, when the left and middle well are initially occupied. The different Hamiltonian models are: extended Hubbard model (green curves), nonstandard Hubbard model (red curves), analytical results from TPA (blue curves). (a,b) Same geometry of Fig.~\ref{fig5} (b) of the main text, i.e. $L=2$, $b=0.5$, $V_0=5$ and $\Omega_0\simeq-0.22$. (c,d) Same geometry of Fig.~\ref{fig5} (c) of the main text, i.e. $L=4$, $b=1$, $V_0=5$ and $\Omega_0\simeq-0.0167$. Interaction strength $V_\delta=3$ and interaction range $\bd/L=2$, so that $\bU\simeq 0.35$, $\Omega_1\simeq 0.0125$ and $\Omega_2\simeq 0.0012$ in (a,b) and $\bU\simeq 0.19$, $\Omega_1\simeq 0.0027$ and $\Omega_2\simeq -6.4 \cdot 10^{-6}$ in (c,d), in arbitrary units.}
\label{fig_appendix_2}
\end{figure*}

In Fig.~\ref{fig_appendix_2}, we illustrate the occupancy probabilities $P_{LM}(t)$ and $P_{LR}(t)$ for the same data as Fig.~\ref{fig5} (b,c) of the main text. Each panel in Fig.~\ref{fig_appendix_2} shows the dynamics obtained from the nonstandard Hubbard model (red curves), the extended Hubbard model (green curves), alongside the analytical results from the TPA (blue curves).

It is evident from Eq.~\ref{app:anP} that when $\bU\gg|\Omega_0|$, single-particle tunneling suppression occurs, as shown in Fig.~\ref{fig_appendix_2} (d). Conversely, for $\bU\sim\Omega_0$, as clearly shown in Fig.~\ref{fig_appendix_2} (b), no suppression is observed. Actually, the nonstandard and extended Hubbard models display significant discrepancies. For instance, analyzing the corresponding probability $P_{LM}(t)$ reveals that while they provide comparable results for the first geometry, see Fig.~\ref{fig_appendix_2} (a), a significant frequency shift arises for the second one, see Fig.~\ref{fig_appendix_2} (c), due to the presence of the term $\Omega_1$, which becomes comparable with $\Omega_0$. More precisely, $\Omega_{1}/|\Omega_0|=0.06$ for the first geometry in Fig.~\ref{fig_appendix_2} (a,b), while $\Omega_{1}/|\Omega_0|=0.16$ for the second one in Fig.~\ref{fig_appendix_2} (c,d).

\begin{figure*}[t]
\includegraphics[width=17.2cm]{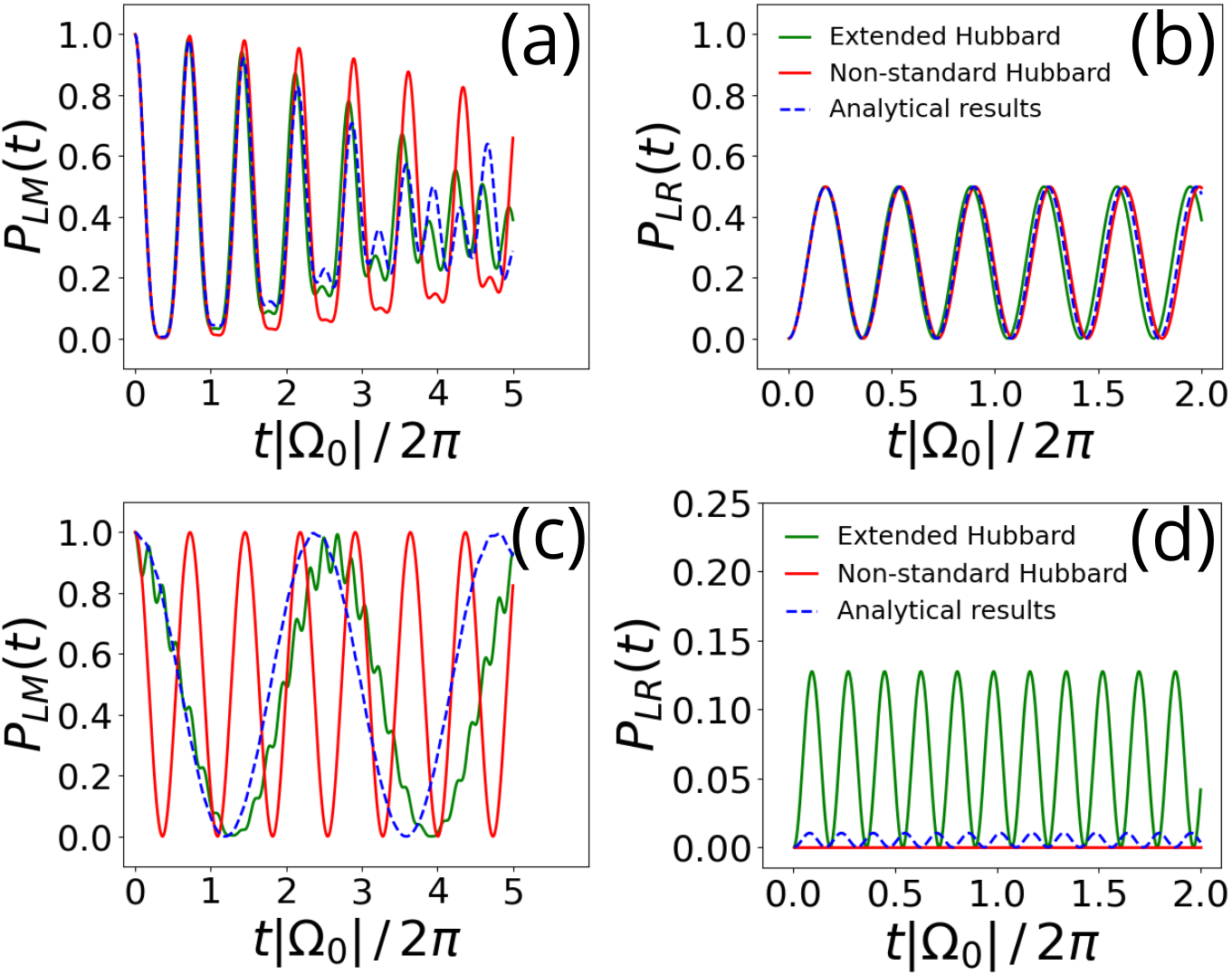}
\caption{Occupancy probabilities $P_{LM}(t)$ (left panels) and $P_{LR}(t)$ (right panels) for two electrons with parallel spins in a triple-well potential, when the left and middle well are initially occupied. The different Hamiltonian models are: extended Hubbard model (green curves), nonstandard Hubbard model (red curves), analytical results from TPA (blue curves). Same geometry of Fig.~\ref{fig6} of the main text, i.e. $L=2$, $b=0.1$, $V_0=1.1$ and $\Omega_0\simeq-0.32$. (a,b) Interaction strength $V_\delta=0.5$ and interaction range $\bd/L=2$, so that $\bU\simeq 0.035$, $\Omega_1\simeq 0.007$ and $\Omega_2\simeq 0.005$. (c,d) Interaction strength $V_\delta=22$ and interaction range $\bd/L=2$, so that $\bU\simeq 1.54$, $\Omega_1\simeq-\Omega_0$ and $\Omega_2 \simeq 0.22$ (complete single-particle tunneling suppression case), in arbitrary units.}
\label{fig_appendix_3}
\end{figure*}

Looking at Fig.~\ref{fig_appendix_2} (b,d), one may have the impression that the predictions concerning the $P_{LR}(t)$ for the two Hubbard models are quite similar. To show that this is not always the case, let us consider the scenario where there is exact cancellation ($\Omega_1=-\Omega_0$), as shown in Fig.~\ref{fig_appendix_3} (c,d) and Fig.~\ref{fig6} of the main text. Looking at Fig.~\ref{fig_appendix_3} (d), one can see that the extended Hubbard model produces oscillations with a significant amplitude of $\approx 0.13$, while the nonstandard Hubbard model yields exactly $P_{LR}(t)=0$. Moreover, the frequency of oscillations of $P_{LM}(t)$ differ notably between the two models, see Fig.~\ref{fig_appendix_3} (c). Specifically, in the extended Hubbard model, the frequency is approximately given by $\Omega_0^2/\bU$ when $\bU \gg \Omega_0$, whereas in the nonstandard Hubbard model, the frequency of oscillations is given by $\Omega_2$.

As the interaction strength decreases, the similarity between the nonstandard and extended Hubbard models is recovered, as shown in Fig.~\ref{fig_appendix_3} (a,b). In such cases, there is no more single-particle tunneling cancellation, and the ratio $\Omega_1/|\Omega_0|$ diminishes significantly (from $1$ to $0.02$).

Finally, comparing the results obtained with the nonstandard Hubbard model to our analytical predictions, we find agreement only when $\Omega_0 \propto \beta \ll 1$, where $\beta$ represents the orbital overlap, see Fig.~\ref{fig_appendix_3} (a,b). Vice versa, discrepancies between the results arise for large $\Omega_0$, due to the neglected terms of order $\Omega_0^2$ in our analytical approximation, see Fig.~\ref{fig_appendix_3} (c,d).

\end{document}